\documentclass[useAMS,usenatbib]{mn2e}

\usepackage{times}
\usepackage{graphicx}
\usepackage{subfigure}

\newcommand{\arone}{148\,GHz}
\newcommand{\artwo}{218\,GHz} 
\newcommand{\arthree}{277\,GHz}
\newcommand{\numberOfSources}{191}
\newcommand{\numberOfRadio}{167}
\newcommand{\numberOfDSFG}{24}
\newcommand{\minimumFlux}{14}
\newcommand{\maximumFlux}{1700\,mJy}

\title[ACT-detected Sources at 148 \& 218 GHz]{The Atacama Cosmology Telescope:  Dusty Star-Forming Galaxies and Active Galactic Nuclei  in the Southern Survey}
\author[D.~W.~Marsden et al.]
{Danica~Marsden$^{1,2}$,
Megan~Gralla$^{3}$,
Tobias~A.~Marriage$^{3}$,
Eric~R.~Switzer$^{4}$,
\newauthor
Bruce~Partridge$^{5}$,
Marcella~Massardi$^{6}$,
Gustavo~Morales$^{7}$,
Graeme~Addison$^{8}$,
\newauthor
J.~Richard~Bond$^{9}$,
Devin~Crichton$^{3}$,
Sudeep~Das$^{10}$,
Mark~Devlin$^{2}$,
Rolando~D\"{u}nner$^{7}$,
\newauthor
Amir~Hajian$^{9}$,
Matt~Hilton$^{11}$,
Adam~Hincks$^{9}$,
John~P.~Hughes$^{12}$,
Kent~Irwin$^{13}$,
\newauthor
Arthur~Kosowsky$^{14}$,
Felipe~Menanteau$^{12}$,
Kavilan~Moodley$^{11}$,
Michael~Niemack$^{15}$,
\newauthor
Lyman~Page$^{16}$,
Erik~D.~Reese$^{2}$,
Benjamin~Schmitt$^{2}$,
Neelima~Sehgal$^{17}$,
\newauthor
Jonathan~Sievers$^{9,16,11}$,
Suzanne~Staggs$^{16}$,
Daniel~Swetz$^{13}$,
Robert~Thornton$^{18}$,
\newauthor
Edward~Wollack$^{4}$\\
\\
$^{1}$Department of Physics, University of California, Santa Barbara, CA 93106, USA\\
$^{2}$Department of Physics and Astronomy, University of Pennsylvania, 209 South 33rd Street, Philadelphia, PA, USA 19104\\
$^{3}$Dept. of Physics and Astronomy, The Johns Hopkins University, 3400 N. Charles St., Baltimore, MD 21218-2686\\
$^{4}$NASA/Goddard Space Flight Center, Greenbelt, MD, USA 20771\\
$^{5}$Department of Physics and Astronomy, Haverford College, 370 Lancaster Avenue, Haverford, PA, USA 19041\\
$^{6}$INAF, Osservatorio Astronomico di Padova, Vicolo dell'Osservatorio 5, I-35122 Padova, Italy\\
$^{7}$Departamento de Astronom{\'{i}}a y Astrof{\'{i}}sica, Facultad de F{\'{i}}sica, Pontificia Universidad Cat\'{o}lica de Chile, Casilla 306, Santiago 22, Chile\\
$^{8}$Department of Physics and Astronomy, University of British Columbia, Vancouver, BC, Canada V6T 1Z4\\
$^{9}$Canadian Institute for Theoretical Astrophysics, University of Toronto, Toronto, ON, Canada M5S 3H8\\
$^{10}$Berkeley Center for Cosmological Physics, LBL and Department of Physics, University of California, Berkeley, CA, USA 94720\\
$^{11}$Astrophysics and Cosmology Research Unit, School of Mathematics, Statistics \& Computer Science, University of KwaZulu-Natal, Durban, 4041, South Africa\\
$^{12}$Department of Physics and Astronomy, Rutgers, The State University of New Jersey, Piscataway, NJ USA 08854-8019\\
$^{13}$NIST Quantum Devices Group, 325 Broadway Mailcode 817.03, Boulder, CO, USA 80305\\
$^{14}$Department of Physics and Astronomy, University of Pittsburgh, Pittsburgh, PA, USA 15260\\
$^{15}$Department of Physics, Cornell University, Ithaca, NY 14853\\
$^{16}$Jadwin Hall, Princeton University, Princeton, NJ, USA 08544\\
$^{17}$Physics and Astronomy Department, Stony Brook University, Stony Brook, NY 11794-3800, USA\\
$^{18}$Department of Physics, West Chester University of Pennsylvania, West Chester, PA, USA 19383}

\begin{document}
\maketitle

\begin{abstract}
We present a catalog of  \numberOfSources\  extragalactic sources detected by the Atacama Cosmology Telescope (ACT) at \arone\ and/or \artwo\ in the 2008 Southern survey.   Flux densities span \minimumFlux\ to \maximumFlux, and we use source spectral indices derived using ACT-only data to divide our sources into two subpopulations: \numberOfRadio\ radio galaxies powered by central active galactic nuclei (AGN), and \numberOfDSFG\ dusty star-forming galaxies (DSFGs). We cross-identify 97\% of our sources (166 of the AGN  and 19 of the DSFGs) with those in currently available catalogs. When combined with flux densities from the Australian Telescope 20\,GHz survey and follow-up observations with the Australia Telescope Compact Array, the synchrotron-dominated population is seen to exhibit a steepening of the slope of the spectral energy distribution from 20 to 148\,GHz, with the trend continuing to 218\,GHz. The ACT dust-dominated source population has a median spectral index, $\alpha_{\mathrm{148-218}}$, of 3.7$^{+0.62}_{-0.86}$, and includes both local galaxies and sources with redshift around 6.  Dusty sources with no counterpart in existing catalogs likely belong to a recently discovered subpopulation of DSFGs lensed by foreground galaxies or galaxy groups.
\end{abstract}


\setcounter{footnote}{0}
\setcounter{table}{0}

\section{Introduction}

\hspace{0.2in} The technologies that enable observations of large numbers of millimeter and submillimeter sources were developed relatively recently.  They open up for study a previously unexplored regime that has the power to reveal the evolution of underlying galaxy populations over cosmic time, and in particular over epochs of intense star formation.  Instruments such as the Submillimeter Common User Bolometer Array \citep[SCUBA, SCUBA-2;][]{holland_1999, holland_2013} operating at 850\,$\mu$m, the Balloon-borne Large Area Submillimeter Telescope \citep[BLAST;][]{pascale_2008} operating at 250, 350 and 500\,$\mu$m, the Large Apex Bolometer Camera \citep[LABOCA;][]{siringo_2009} operating at 870\,$\mu$m, and the AzTEC millimeter wavelength camera \citep{wilson_2008} operating at 1.1 and 2.1\,mm have mapped up to 10\,deg$^2$ of the sky, but greater coverage has been limited by the large amount of integration time required to conduct blind surveys to significant cosmological depth.  Increased sky coverage and sensitivity motivated the construction of space-based observatories such as {\it Spitzer} \citep{werner_2004} and {\it Herschel} \citep{pilbratt_2010}, operating in the wavelength regime 3--500\,$\mu$m and covering up to tens of square degrees.  At longer wavelengths, the Wilkinson Microwave Anisotropy Probe \citep[{\it WMAP};][]{wright_2009} covered the whole sky at frequencies up to 94\,GHz ($\lambda$\,=\,3.2\,mm), but observations of unresolved extragalactic sources were only complete above 2\,Jy due to its large beam size (roughly 13$\arcmin$ at 94\,GHz).  The {\it Planck} space telescope contains a high frequency instrument observing at frequencies spanning 100--857\,GHz \citep{lamarre_2010}.  At 143 and 217\,GHz (2.1 and 1.4\,mm), the {\it Planck} beam sizes are approximately 7\arcmin\ and 5\arcmin\ respectively, allowing for a source catalog that is complete down to flux densities of about 1\,Jy.  Large-area ($>$\,100\,deg$^2$) ground-based radio surveys probe with higher resolution than {\it WMAP} or {\it Planck}, but only as high in frequency as 20\,GHz \citep[e.g., the Australia Telescope 20\,GHz survey;][]{murphy_2010}.  Thus there is a niche for millimeter wavelength large-scale mapping experiments with spatial resolution superior to that of the space-based observatories.

One such experiment is the Atacama Cosmology Telescope \citep[ACT;][]{swetz_2011}.  The ACT collaboration released Cosmic Microwave Backgound (CMB) temperature anisotropy maps made with arcminute resolution from its 2008 observing season at 148 and 218\,GHz \citep{dunner_2013}.  These maps also contain galaxies that are luminous at millimeter wavelengths.  Measurements of the millimeter fluxes of a large sample of sources have great potential to discriminate among source population models and reveal new source populations.  For example, recent millimeter surveys \citep[e.g.,][]{vieira_2010, marriage_2011a, planck7_2012} have yielded source counts that led to updated source models \citep{tucci_2011} and the discovery of a new class of high-flux dusty galaxies \citep[this publication;][]{vieira_2010,negrello_2010}.  

ACT-detected sources at flux densities greater than 20\,mJy are predominantly blazars that have synchrotron-dominated spectral energy distributions (SEDs). These galaxies are powered by active galactic nuclei (AGN), central black holes that accrete nearby matter in a process that produces time-variable jets out of the plane of the accretion disk; in the case of blazars, the jet direction lies close to our line of sight.  Ejected ionized particles spiraling around magnetic field lines in the jets create the observed synchrotron emission.  The study of blazar SEDs and source counts provides information about AGN physics \citep[e.g.,][]{dezotti_2010,tucci_2011, planck15_2011}.

The second population identified at ACT wavelengths is comprised of infrared-luminous, dusty star-forming galaxies (DSFGs), which exhibit modified blackbody emission at sub-millimeter to millimeter wavelengths, diminishing toward longer wavelengths.  The observed SED is dominated by thermal emission from dust grains that have been heated by the prodigious optical and ultraviolet flux produced by newly formed stars \citep[e.g.,][and references therein]{draine_2003}.  A fraction of these galaxies may have a significant non-thermal contribution from AGN at their cores, in addition to the thermal dust emission.

\begin{figure*}
  \begin{center}
 \subfigure{\includegraphics[scale=.67, clip, trim=14mm 35mm 24mm 22mm]{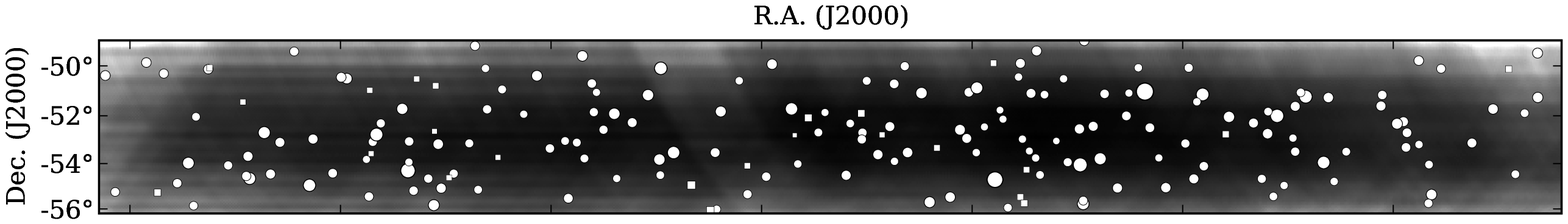}}
  \subfigure{\includegraphics[scale=.67, clip, trim=14mm 9mm 24mm 42mm]{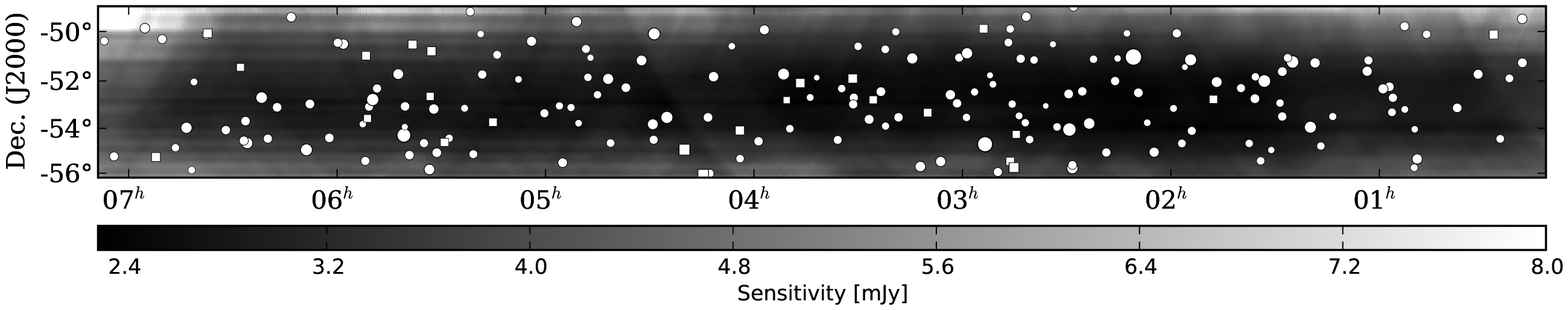}}
  \end{center}
  \caption[Sensitivity maps with source detections at 148 and 218\,GHz.]{Sensitivity maps with source detections at 148\,GHz (top) and 218\,GHz (bottom), showing the most uniform 455\,deg$^2$ patch of the \emph{Southern strip} with the greatest depth of coverage.  It lies between right ascension $00^{\mathrm h}12^{\rm m}$ and $07^{\mathrm h}07^{\rm m}$, and declination $-56\degr11\arcmin$ and $-49\degr00\arcmin$.  The deepest data correspond to an exposure time of 23.5\,minutes per square arcminute and a 1$\sigma$ sensitivity of 2.34\,mJy at 148\,GHz and 3.66\,mJy at 218\,GHz. White circles or squares mark the locations of ACT sources with a size proportional to the log of the associated source flux density.  Circles denote sources designated as AGN, and squares denote sources with spectra indicative of DSFGs.  Toward the edges of the map, the variation in local noise properties due to uneven coverage is more apparent.}
  \label{fig:coverage}
\end{figure*}

Some of these DSFGs are local galaxies ($z$\,$\ll$\,1), cross-identified with Infrared Astronomy Satellite (IRAS) sources \citep{devereux_1990}.  The IRAS wavebands ($\lambda$\,=\,12--100\,$\mu$m) are relatively insensitive to dust with temperatures below 30\,K, a significant and largely unexplored component of many nearby galaxies \citep{planck17_2011}.  Recent results from {\it Herschel} \citep{amblard_2010} and BLAST \citep{wiebe_2009} have begun to extend our picture of the cold dust in galaxies, but millimeter-wavelength experiments can, through probing dust in nearby galaxies, contribute to establishing a well-calibrated dust SED for typical, low-redshift DSFGs.  This in turn has cosmological ramifications, as current analyses of the first generations of stars and galaxies that fuel the cosmic infrared background (CIB) rely on understanding and extrapolating from dust SED templates.

High-redshift DSFGs were observed by SCUBA in the first systematic survey of these sources, which create a significant fraction of the CIB emission \citep{blain_2002}.  Selected at 850\,$\mu$m, these sources came to be known as submillimeter galaxies (SMGs).  This new population of galaxies subsequently became a focus of observations \citep[e.g.,][]{weil_2009, viero_2009, austermann_2010} and stacking analyses that resolved more of the CIB into emission from discrete, dusty star-forming galaxies \citep[e.g.,][]{dole_2006, devlin_2009}.  Though most SMGs will be undetectable by current millimeter survey instruments, a new population of sources that are significantly brighter and rarer than the submillimeter-selected SMGs and that similarly exhibit dust-dominated spectral indices has been identified at millimeter wavelengths \citep{vieira_2010}.  They do not have counterparts in the IRAS catalog, indicating that they are not members of the standard, local, ultraluminous infrared galaxy (ULIRG) population.  

These sources have recently been shown to be a new, higher redshift ($z$\,$>$\,3) subpopulation of the progenitor galaxy background, brought to the fore in millimeter wavelength surveys because they are lensed by foreground galaxies or galaxy groups \citep{negrello_2007, lima_2010, negrello_2010, vieira_2013,hezaveh_2013}.  Such objects are extremely rare \citep[at submillimeter wavelengths, for example, see][]{rex_2010, lupu_2012}, but wide area surveys will find more.  The South Pole Telescope \citep[SPT;][]{carlstrom_2011} has reported significant numbers of these sources.  This population provides an avenue for follow-up research to study the details not only of lensed SMGs, but also of the lens systems \citep[e.g.,][]{ikarashi_2010, scott_2011, lupu_2012, weil_2013}.  

In this paper, we report for the first time the discovery of DSFGs in the ACT data.  This is the first multifrequency analysis of the ACT sources and the second report on ACT extragalactic sources.  \cite{marriage_2011a}, hereafter M11, presented a catalog of sources at \arone\ only from the 455\,deg$^2$ of the 2008 ACT Southern survey with the best uniformity and coverage.  Here we extend those results to include 218\,GHz (and updated 148\,GHz) flux densities for the same 2008 ACT Southern survey region from the new \cite[][hereafter D13]{dunner_2013} data release maps.  Gralla et al., in prep. will present sources detected in the ACT \emph{Equatorial strip}.  Mapmaking of the ACT 277\,GHz dataset is currently under way. 

The layout of the paper is as follows.  Section~\ref{sec:data} describes the 2008 season ACT observations and the reduction of raw data into maps, as well as follow-up observations made with the Australian Telescope Compact Array (ATCA).  Section~\ref{sec:sourceExtraction} details our method of source extraction.  The source catalog, including its astrometric and flux density accuracy, and its estimated completeness and purity, is discussed in Section~\ref{sec:catalog}. Section~\ref{subsec:others} compares our catalog with currently available datasets; Section~\ref{subsec:counts} gives the source number counts.  Trends observed in the spectral indices of our source populations are analyzed in Section~\ref{subsec:spectra}.  We conclude in Section~\ref{sec:conc} with a summary of our results.

\section{Observations and Data}
\label{sec:data}

\subsection{ACT Observations}

\hspace{0.2in} The ACT experiment \citep{swetz_2011} is situated on the slopes of Cerro Toco in the Atacama Desert of Chile\footnote{22.9586$^{\circ}$ south latitude, 67.7875$^{\circ}$ west longitude.} at an elevation of 5,190\,m. ACT's latitude gives access to both the northern and southern celestial hemispheres.  Observations occurred simultaneously in three frequency bands, at \arone\ (2.0\,mm), \artwo\ (1.4\,mm) and \arthree\ (1.1\,mm) with angular resolutions of roughly 1.4\arcmin\,, 1.0\arcmin\ and 0.9\arcmin\ respectively. Observations of Saturn were used to determine beam profiles and pointing \citep{hincks_2010, hasselfield_2013}. From 2007 to 2010, ACT targeted two survey regions: the \emph{Southern strip} centered around $\delta$\,=\,$-52.5^{\circ}$ and the \emph{Equatorial strip} centered around $\delta$\,=\,$0^{\circ}$. Further information about the ACT observations can be found in D13.

\subsection{ACT Data}
\label{subsec:actdata}

\hspace{0.2in} The reduction of raw ACT data into maps is detailed in D13; we will briefly review that process here.  Each ACT detector array (one per frequency band) is composed of 1,024 detectors, with each detector timestream first analyzed and then kept or rejected based on multiple criteria, such as telescope operation, weather conditions, cosmic ray hits, or other interference.  Approximately 800\,(700)\,hours of data from the 2008 \emph{Southern strip} for \arone\ (\artwo) remain after these cuts.

Maximum likelihood maps of pixels 30\arcsec\ on a side are produced from the timestream data.  From an initial estimate of the maps, source profiles are fit for.  We then subtract the timestream models for the point sources from the data and re-map.  This prevents point source power from being aliased into the map, and improves the final flux density estimates for S/N\,$>$\,5 sources (quantified below).  Similarly, an estimate of the CMB signal is also removed.  The mapping equation \citep{tegmark_1997} is then solved iteratively using the preconditioned conjugate gradient method.  Finally, the signals that had been subtracted are added back to the map.  We conducted simulations in which a signal corresponding to sources of known flux were injected into  raw ACT detector time streams before mapping.  The new mapping procedure with ``source subtraction'' described briefly here and more thoroughly in D13 (Section 11.4) corrects a 3-5\% downward bias in the recovered flux densities. 

The measured source flux densities have converged on a single value by the 25$^{th}$ iteration of the maps used in our study.  For \arone, the D13 release map is from iteration 1000 of the map-maker, and we use this map both for convenience and in order that this study be exactly reproducible using public data.  For our study of source flux densities at 218\,GHz, we use iteration 200, which is more than adequate.  Between the map iterations used in this study, the fluxes of the sources change less than $1\%$.  One may ask whether the value to which the measured source flux density has converged is accurate. Based on end-to-end simulations in which mock sources are injected into the ACT time streams, we estimate this accuracy at 3\%. 

The \arone\ map is calibrated in temperature through cross comparison of spectra over the range 400\,$<$\,$\ell$\,$<$\,1000 with WMAP as in \cite{hajian_2010} with an uncertainty of 2\%.  This calibration is transferred to the \artwo\ map (2.4\% uncertainty) through cross-calibration with \arone.  \cite{hasselfield_2013}, hereafter H13, used this calibration in measurements of the temperatures of Uranus and Saturn. The resulting planet temperature estimates were consistent with previous analyses \citep{griffin_orton_1993,goldin_1997}.  To summarize the results of H13, both previous analyses adopted a flux standard based on a model of Mars \citep{wright_1976, ulich_1981} with 5\% systematic errors. The H13 estimate for the temperature of Uranus and Saturn was 5\% below those of the previous analyses, a deficit which was also seen in the WMAP analysis of \cite{weiland_2011} at 94\,GHz and is most likely due to systematic errors in the Mars standard.  The agreement between the H13 analysis, calibrated on extended CMB emission, and the previous analyses, calibrated on Mars, lend confidence that the ACT WMAP-based calibration at large angular scales will also apply to the calibration of sources at small angular scales. 

Photometry based on matched filtering (Section \ref{sec:sourceExtraction}) relies on an accurate estimate of the instrument beam shape. The impact of beam error on photometry depends on the details of the filter (Equation \ref{eq:filt}), which in turn depend on both the beam window function and the angular power spectrum of the background and noise. We have found that the error in the solid angle of the beam is a conservative proxy for the photometric error due to beam uncertainty, and we use this simpler quantity (solid angle uncertainty) in the beam-related photometry error estimates described here. Due to uncertainties in the beam measurements from Saturn we assign $1\%$ photometric errors for both bands. The profiles of bright point sources in the survey maps suggest that the effective beam for the survey is slightly broader than that measured from Saturn. The survey maps consist of overlapping observations taken over the entire season, and the observed broadening may be attributed to a night-by-night jitter in the telescope pointing with an rms of 5\,$\pm$\,1\arcsec \citep{hasselfield_2013}. The uncertainty in jitter correction corresponds to $1\%$ (\arone) and $2\%$ (\artwo) photometric uncertainties, which are correlated between bands.  

Finally, for our photometric uncertainty budget we account for the fact that AGN have a lower effective frequency band center (and thus slightly broader beam) than Saturn, and DSFGs have a higher effective frequency band center \citep[and thus narrower beam;][Table 4]{swetz_2011}. We choose effective \arone\ and \artwo\ frequency centers corresponding to halfway between a steep spectrum AGN and a DSFG: 148.65\,GHz and 218.6\,GHz. This choice introduces a photometric bias of less than 1.5\% at 148\,GHz and 1.1\% at 218\,GHz, which is positive for DSFGs and negative for steep spectrum AGN.  We fold this  photometric bias from the source spectrum in with uncertainties due to mapping (3\%), WMAP calibration (2\%, 2.4\%) and beam shape (1.4\%, 2.2\%) to obtain an overall flux density calibration uncertainty of 4.1\% at \arone\ and 4.6\% at \artwo. As shown in Section \ref{subsec:others}, the ACT flux densities agree with independent measurements to within this margin of error.

For this study, we have used the most uniform 455\,deg$^2$ of the 2008 ACT \emph{Southern strip} at 148 and 218\,GHz. This region, shown in Figure \ref{fig:coverage}, spans declination -56.2$^\circ$\,$<$\,$\delta$\,$<$\,-49.0$^\circ$ and right ascension $00^{\mathrm h}12^{\rm m}$\,$<$\,$\alpha$\,$<$\,$07^{\mathrm h}07^{\rm m}$.  At 148\,GHz we use the data publicly released with D13\footnote{http://lambda.gsfc.nasa.gov/product/act/}. Typical white noise levels in this region of the map are 30\,$\mu$K-arcminute at \arone\ and 50\,$\mu$K-arcminute for \artwo. As described in Section \ref{sec:sourceExtraction}, when matched-filtered with the ACT beam, this white noise level results in a 1$\sigma$ point source flux density sensitivity in the best covered regions of $\sigma_0$\,=\,2.34\,mJy at \arone\ and 3.66\,mJy at \artwo.  The sensitivity levels in Figure \ref{fig:coverage} are proportional to the square root of the number of observations at that map location, $N_{\mathrm{obs}}$, with one observation per 0.005 seconds.  Then the sensitivity level in a given portion of the map is $\sigma$\,=\,$\sigma_0 \sqrt{N_{\mathrm{obs, max}}/N_{\mathrm{obs}}}$.

\subsection{ATCA Observations}
\label{subsec:atcaobs}

\hspace{0.2in} \cite{marriage_2011a} found that the sample of ACT \arone\,-detected sources cross-identified with the Australian Telescope 20\,GHz survey \citep[AT20G;][]{murphy_2010} is dominated by sources with peaked or falling SEDs using flux densities measured at 5, 20 and 148\,GHz.  The study also confirmed the findings of the AT20G study \citep{murphy_2010, massardi_2011a}, namely that this population of radio sources is characterized, on average, by spectral steepening between 20--30\,GHz and 148\,GHz.  However, this sample from M11 was incomplete, biased in a way that favored sources with negative spectral indices between 20 and 148\,GHz due to the AT20G survey completeness level of 78\% above 50\,mJy at 20\,GHz.  ACT-selected sources with 148\,GHz flux densities less than 50\,mJy and flat or rising spectra may not have been detected by AT20G.  Therefore, in order to complete the M11 20--148\,GHz spectral study, a targeted set of measurements of flux densities at 20\,GHz was made for the M11 sources that were not identified with any source in the AT20G catalog within a 30\arcsec\ search radius. 

Scheduling and weather constraints permitted us to observe 41 of these 48 sources with the 6\,$\times$\,22\,m antenna array of ATCA\footnote{http://www.narrabri.atnf.csiro.au/} over 6.5\,hours at 20\,GHz on November 10, 2010, when the array was in its East-West 750A configuration, with 15 baselines ranging from 77 to 3750\,m.  The primary beam of each array telescope at 20\,GHz is 2.3\arcmin\,, with a resolution of 0.5\arcsec\, possible with the full array in this configuration.  The synchrotron-dominated spectra of these sources, as revealed in M11, indicated that they were compact (AGN), and would be unresolved.  For these observations, the two 2\,GHz-wide frequency bands of the new ATCA Compact Array Broadband Backend \citep[CABB;][]{wilson_2011} digital array correlators were set to be adjacent by centering them at 19\,GHz and 21\,GHz. The reduced average flux densities over the whole bandwidth of the correlator corresponds to the ``20\,GHz'' flux density.  Good weather conditions prevailed throughout the observations, from 07:35:24.9\,UT through 14:11:54.9\,UT.  A few data blocks were flagged and removed in order to minimize noise and any spurious effects.

The primary flux calibrator used for these observations was PKS B1934-638, and the bandpass calibrator used was PKS B1921-293.  The target sources were expected to have flux densities at or just below the AT20G survey limit of 40\,mJy.  Each target source was observed once for 1.5\,minutes for an rms noise level of $\leq$\,0.15\,mJy/beam. Target observations were interleaved with observations of four secondary calibrators for pointing and phase corrections, chosen to lie close to the targets.  

\subsection{ATCA Data}

\hspace{0.2in} The ATCA follow-up data was reduced using a fully automated, custom, shell script pipeline based on tasks from the MIRIAD aperture synthesis reduction package \citep{sault_1995}.  An initial inspection of the data was performed to identify contamination or any problems in the data acquisition. Automatic procedures were used to identify and flag data for each frequency band affected by shadowing or known radio contamination, resulting in less than 1\% of the band being flagged and cut.  The pipeline then generated the calibration solutions for bandpass, flux density amplitude and phase based on the calibrators.  After checking that results were consistent between the two frequency bands, we merged the calibrated target visibilities before extracting the flux densities to improve sensitivity. 

Analyses of the AT20G survey \citep{murphy_2010} and of other ATCA projects \citep{massardi_2011b, bonavera_2011} have shown that the triple correlation technique \citep{thompson_1986} can be effectively used to determine source flux densities down to tenth of mJy scales for point-like sources unaffected by poor phase stability.  This technique is effective for point sources even if they are not in the phase center and in the case of low S/N. 

We manually inspected the 20\,GHz deconvolved images, and wherever a source was clearly identifiable we measured its flux density and flux density error at its peak in the image and the rms of the image pixels in a region unaffected by the source emission. For sources with high S/N, deconvolution techniques were able to reconstruct images, despite the poor $uv$-coverage of our observations.  In the few cases where the source appeared extended with respect to the synthesized beam we estimated the flux density by integrating over the source area.  In some cases, the image showed a confused field with multiple peaks, primarily due to sidelobes; for these cases we extracted a flux density estimate using the triple correlation technique.  The flux density errors were obtained from the rms of the visibilities of the V Stokes parameter.  This assumes that these objects have negligible circular polarization, which is an overestimate in the cases where the source is not in the phase center.  The same statistic (V Stokes rms) was used to indicate the noise level reached for fields where no source was identified at 20\,GHz.  Of the 41 ACT sources observed in the follow-up campaign described here, all had a measured flux density at 20\,GHz. The final results of this study are included as part of Table 4 with asterisks indicating sources that were part of the follow-up ATCA sample.

\section{ACT Source Extraction}
\label{sec:sourceExtraction}

\hspace{0.2in} We use the same matched filtering method as described in M11 to produce an estimate of the amplitude of a source at the location of the source center. In this section we summarize the method with an emphasis on aspects for which this analysis differs from M11.  

Initially, the data are weighted by the square root of the number of observations per pixel.  This method results in a map with an approximately constant white noise level and a natural apodization as the number of observations falls off toward the edge of the survey region.  Using a new procedure, we filter the entire D13 \arone\ release map and corresponding \artwo\ map, both of which extend beyond the deep 455\,$\deg^2$ area of this study, in order to eliminate any potential edge effects. We then use only the 455\,$\deg^2$ region for source identification.

\begin{figure}
  \begin{center}
  \subfigure{\includegraphics[scale=0.48, clip, trim=13mm 3mm 8mm 13mm]{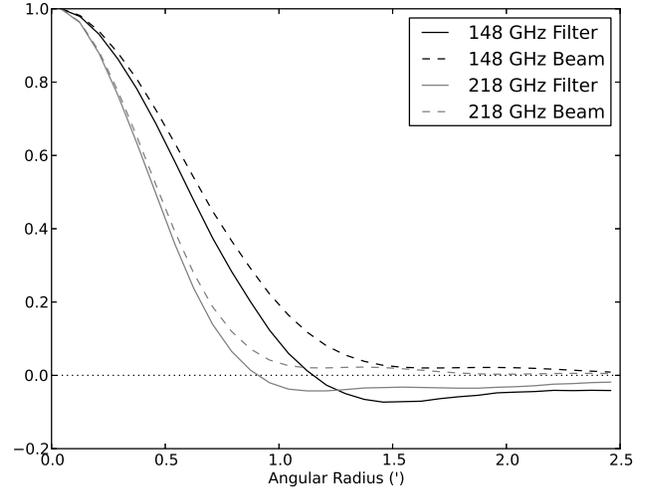}}
  \end{center}
  \caption[Matched filters and instrument beams.]{Matched filters (solid lines) are band pass filters that smooth the map on the scale of the beam (dashed lines) and remove large scale structure associated with the CMB, other astrophysical signals (e.g., the Sunyaev-Zel'dovich effect) and residual contamination from atmospheric brightness fluctuations.  The beam and filter functions are plotted with unit normalization at their peak ($\theta$\,=\,0). The matched filters were binned in radius  to better show the relevant angular scales. In the analysis, we use two dimensional filters to capture the anisotropic character of the noise. }
  \label{fig:filters}
\end{figure}

The following filter is applied to the Fourier transform of the map: 
\begin{equation}
\Phi(\mathbf{k})=\frac{ F_{k_0,k_x}(\mathbf{k})\tilde{B}^*(\mathbf{k})|\,\tilde{T}_{\rm other}(\mathbf{k})\,|^{-2}}{\int \tilde{B}^*(\mathbf{k'})F_{k_0,k_x}(\mathbf{k'})|\,\tilde{T}_{\rm other}(\mathbf{k'})\,|^{-2}\tilde{B}(\mathbf{k'})\,d\mathbf{k'}} ,
\label{eq:filt}
\end{equation}
where $\mathbf{k}$\,=\,$(k_x,k_y)$ is the angular wave number, and $x$ and $y$ refer to the right ascension and declination directions. ${B}(\mathbf{k})$ is the Fourier transform of the ``effective'' instrument beam. As described in Section \ref{sec:data}, the ``instantaneous'' beam is derived from planet observations \citep{hasselfield_2013}. The ``effective beam'' in the 2008 survey map is broadened by imperfect telescope repointing with mean deviation $\sigma_\theta$\,=\,5$\arcsec$. This broadening is included in ${B}(\mathbf{k})$ by multiplying the instantaneous beam transform  (released with D13 for \arone) by $\exp(-\ell^2\sigma_\theta^2/2)$.  $\tilde{T}_{\rm other}$ is the Fourier transform of all components of the data besides point sources (i.e., atmospheric or detector noise, CMB, etc.).  The function $F_{k_0,k_x}(\mathbf{k})$ is a high-pass filter that removes undersampled large scale modes below $k_0$\,=\,1000 and modes with $\mid$\,$k_x$\,$\mid$\,$<$\,100, which are contaminated in a fraction of ACT data by telescope scan-synchronous noise. While ${B}(\mathbf{k})$ is well approximated as azimuthally symmetric, we retain the full two dimensional power spectrum $|\,\tilde{T}_{\rm other}(\mathbf{k'})\,|^{2}$ to downweight anisotropic noise in the maps. The azimuthally-binned 148 and \artwo\ real space filters and associated beam profiles are shown in Figure \ref{fig:filters}.

The power spectrum  $|\,\tilde{T}_{\rm other}(\mathbf{k'})\,|^{2}$ used in this analysis was constructed in a different manner from M11: we used the power spectrum of the data itself instead of the average of difference (noise) maps and models for the CMB and other contaminating sky emission. This estimate is robust since the total power from the extragalactic source signal is low compared to the CMB, atmospheric noise, and white noise.  To avoid a noisy estimate of the power spectrum, we smooth the power spectrum $|\,\tilde{T}_{\rm other}(\mathbf{k'})\,|^{2}$ with a Gaussian. The exact formulation of the smoothing does not significantly change the resulting filtered map. 

Applying the matched filter can cause ringing in the maps around the very brightest sources, which impacts source extraction around other, low S/N, sources.  Therefore, we identify sources with S/N\,$>$\,50 in an initial application of the filter, and mask them in the maps. Sources with S/N\,$<$\,50 are then extracted by match filtering these masked maps.  Once filtered, groups of map pixels with S/N\,$>$\,4.8 are identified as candidate sources.

Finite map pixel size affects the measured flux densities, since it is rare that a source falls exactly in the center of a pixel, leading to a systematic negative bias and increased scatter.  We increase the map pixel resolution by a factor of 16 in a region 0.03$^{\circ}$ on a side around each source through Fourier space zero-padding \citep[e.g.,][]{press_2007} to allow for a more precise determination of the source peak location, and therefore its flux density.  Properly centering the detection has the effect of boosting the S/N of typical S/N\,=\,4.8 sources to $\ge$\,5.  Lastly, to account for the convolution of sources with the map pixel, which acts as a low pass filter (the square pixels become a sinc function in Fourier space), we deconvolve the map pixel window function in the higher resolution map. We find that this method reduces systematic errors associated with pixelization to below 1\%.  Purity tests (Section \ref{subsec:purityAndCompleteness}) reveal a significant number of false detections below S/N of 5.0, due to local noise and striping artifacts.  Therefore we impose a S/N\,=\,5.0 threshold on sources detected in each map assuming no prior knowledge of source location from the other frequency.   

\begin{figure}
\begin{center}
  \subfigure{\includegraphics[scale=.48, clip, trim=7mm 8mm 17mm 15mm]{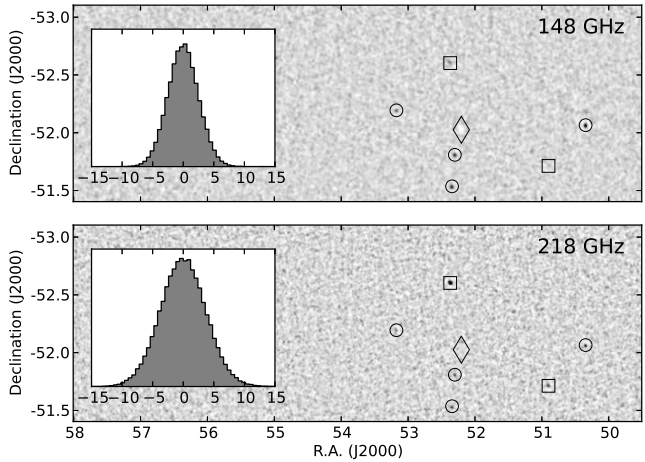}}
\caption{  Filtered 148\,GHz (top) and 218\,GHz (bottom) submaps. The data have been match-filtered such that the grey-scale is in units of flux density (mJy) with white (black) corresponding to -10\,mJy (30\,mJy). Insets show the flux density distribution across the data as a grey histogram that has a standard deviation of 2.57\,mJy (3.78\,mJy) for 148\,GHz (218\,GHz).  Several sources, both synchrotron and dust dominated, marked as circle and square outlines respectively, are apparent as black beam-sized flux excesses. The white extended object in the 148\,GHz map, marked by a diamond,  is the Sunyaev Zel'dovich effect decrement from Abell 3128 NE (ACT-CL J0330-5228).}
\label{fig:map}
\end{center}
\end{figure}

Once a S/N\,$>$\,5.0 catalog has been generated for each of the filtered \arone\ and \artwo\ maps, we look for sources that have been identified at one frequency but not the other.  The flux density for the source at the second frequency is then measured from the filtered map for the second frequency.  We use the same prescription as for the flux measurement method just described, using a submap at the second frequency centered on the location of the detection as determined from the map at the first frequency.

\section{The 148 and \artwo\ Catalog}
\label{sec:catalog}

\hspace{0.2in} The ACT-detected source catalog is given in Table~4. We find 169 sources selected at \arone\ with S/N\,$>$\,5, spanning two decades in flux density, from \minimumFlux\ to \maximumFlux.  The \artwo\ map independently yielded 133 sources with S/N\,$>$\,5.  The combination of these two independent source lists gives a total count of \numberOfSources, with 110 galaxies detected with S/N\,$>$\,5 at both frequencies.  

The catalog provides the IAU name, celestial coordinates (J2000), S/N and flux density estimation of each 148 and 218\,GHz ACT-detected source. Raw flux densities are estimated directly from the map as described in Section \ref{sec:sourceExtraction}.  Deboosted flux densities as derived according to Section \ref{subsec:deboosting} are given with associated 68\% confidence intervals.  The raw and deboosted spectral indices, $\alpha_{148-218}$, between 148 and 218\,GHz for each source, are provided.  The classification of sources as dust or synchrotron-dominated is based on the $\alpha_{148-218}$ spectral index criterion described in Section \ref{subsec:deboosting}.  If the source was cross-identified with an AT20G catalog source (see Section \ref{subsec:others}), the AT20G source ID is given.  If, instead, the 20\,GHz source flux density was measured during the November 2010 follow-up campaign (Section \ref{subsec:atcaobs}), the 20\,GHz ID name has an asterisk next to it. Sources not cross-identified with one of the catalogs listed in Section~\ref{subsec:others} are marked with a ``d'' superscript.   

Correlating 148\,GHz flux densities between this catalog and the catalog given in M11, we find an average agreement at the 2\% level, with larger scatter for individual sources. This level of consistency is expected given changes in calibration, map-making, beam profile estimates, and deboosting procedure made for this updated study.

The reported flux densities are the average over approximately two months of observation for each source, many of which are AGN-driven radio galaxies and thus likely to have varied in that time. For example, of the three bright sources cross-identified with {\it Planck}, in the year between ACT and {\it Planck} observations one source varied by 20\% in flux density whereas another did not vary at all within errors.  However, our simultaneous multifrequency observations allow for a consistent internal spectral characterization between ACT bands.    In a future study, multiple years of data will be used to quantify the effects of variability on individual source spectra. 

The following sections provide details and context for the catalog values.

\subsection{Astrometric Accuracy}
\label{sec:astrometry}

\hspace{0.2in} Radio interferometers can achieve very precise positional accuracy for sources, so ACT-selected sources cross-identified with a robust radio catalog give a good measure of the positional accuracy of the ACT source detections.  The AT20G catalog covers the Southern sky, and through pointing checks against Very Long Baseline Interferometer (VLBI) measurements of International Celestial Reference Frame calibrators, the positional accuracy of AT20G is shown to be accurate to better than 1\arcsec\ \citep{murphy_2010}.  We exclude nearby extended/resolved sources from this analysis, determined from cross-identification of our sources with currently available catalogs, as distant point-like sources will present a clearer picture of the overall accuracy of our pointing.   

\begin{figure}
  \begin{center}
  \includegraphics[scale=0.85, clip, trim=1mm 10mm 10mm 18mm]{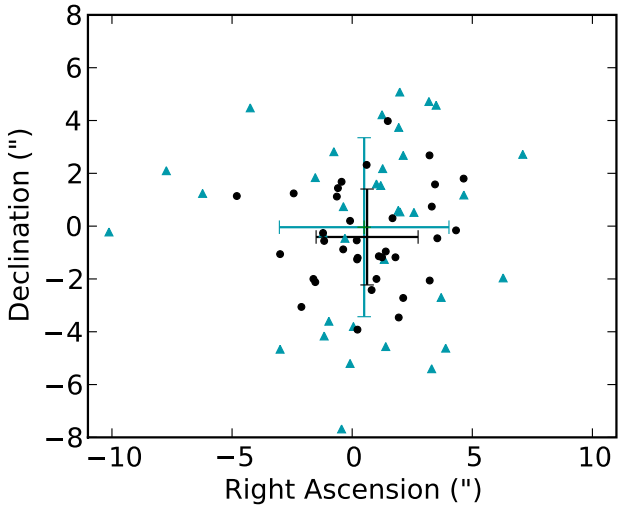}
  \end{center}
  \caption[Astrometric accuracy of the ACT source detections.]{ Astrometric accuracy of the ACT source detections. The 34 filled black circles are the positional offsets of ACT sources with S/N$_{148}$\,$>$\,16 from counterparts in the AT20G catalog as calculated from the 148\,GHz map. The turquoise triangles are for the same sources, but using positions derived from the 218\,GHz map, which has more noise.  The large errorbars are centered on the mean offsets, and extend as far as the rms of the offsets.  See Table~\ref{tab:astrometry}. }
  \label{fig:pointing}
\end{figure}

We compared the positions of the 34 ACT sources with S/N$_{148}$\,$>$\,16 to positions of associated sources in the AT20G catalog using a search radius of half the ACT 148\,GHz beam (0.7\arcmin).  Figure \ref{fig:pointing} shows the offsets in location between the AT20G right ascension (RA) and declination (Dec) and the ACT-derived positions for the sources as they were detected in the 148\,GHz (black points) and 218\,GHz (blue points) maps.  The results are also summarized in Table \ref{tab:astrometry}.  The large overlaid errorbars are centered on the mean offsets, and extend as far as the rms of the offsets.   For sources with lower S/N, the ACT location rms with respect to AT20G position becomes inflated by the effect of noise in the maps.  Therefore the catalog uses coordinates derived from the 148\,GHz map since there is a higher level of noise present in the 218\,GHz map.

\begin{table}
\caption{Astrometric Pointing Accuracy}
\label{tab:astrometry}
\begin{center}
\begin{tabular}{lcc}
\hline
\hline
 & 148\,GHz  & 218\,GHz  \\
\hline
Mean RA Offset ($\arcsec$) & 0.6\,$\pm$\,0.4  &  0.5\,$\pm$\,0.6\\
Mean Dec Offset ($\arcsec$)& $-0.4$\,$\pm$\,0.3  & 0.0\,$\pm$\,0.6 \\
\hline
RMS in RA Offset ($\arcsec$)&  2.1  & 3.5 \\
RMS in Dec Offset ($\arcsec$)& 1.8  & 3.4 \\
\hline
\end{tabular}
\end{center}
\end{table}

\subsection{Flux Density Accuracy}
\label{subsec:deboosting}

\hspace{0.2in} Flux densities derived from the ACT maps have systematic uncertainties arising from five effects: the overall calibration uncertainty,  the mapmaker, errors in the assumed source profile, error in the assumed source spectrum, and flux boosting of lower significance candidates.  The first four potential sources of flux density error were discussed in Section \ref{sec:data}.   Here we will discuss the last source of flux density error.

The differential counts of the sources selected in our sample fall steeply with increasing flux (Table~\ref{tab:stats}, Figure~\ref{fig:dNdS}). With no prior information about the source flux, the most likely scenario is that the measured flux is the sum of a dimmer intrinsic flux and a positive noise fluctuation. We use the two-band Bayesian method developed in \cite{crawford_2010}, and report the 16, 50, and 84 percentiles ($68\%$ enclosed, equivalent to $1\sigma$) of the posterior flux and spectral index distributions. For the source count priors in this calculation we use the sum of the models of \cite{dezotti_2010} for radio sources (using the Tucci et al. 2011 model results in flux differences of $<$\,0.03$\sigma$) and \cite{bethermin_2011} for dust-dominated sources. Following \cite{vieira_2010}, we take a flat prior on the spectral index between $-3$ and $5$, consistent with the expected range for our populations. The two-band likelihood includes negligible correlation between bands and is consistent with background astrophysical emission rather than correlated atmospheric emission. 

The source populations in 148 and 218\,GHz naturally split into sources having their emission dominated by synchrotron (centered on $\alpha$\,=\,-0.6) or thermal dust (centered on $\alpha$\,=\,3.7; see Figure~\ref{fig:spectralIndices}, bottom panel). We use the threshold spectral index $\alpha$\,=\,1.66 \citep{vieira_2010} to divide these populations in terms of their posterior spectral index populations, with $P(\alpha$\,$>$\,1.66)\,$>$\,0.5 classified as dusty and $P(\alpha$\,$>$\,1.66)\,$<$\,0.5 as synchrotron dominated. The classification is robust within $\alpha$\,$\pm$\,0.5 of this threshold.  While calibration uncertainty is included in all quoted fluxes, we conservatively ignore inter-band correlation of calibration errors when flux deboosting to avoid making assumptions about the correlation of the pointing jitter, chromatic effects, and mapping/flux recovery errors (Section~\ref{subsec:actdata}). This moderately boosts the error on the spectral index toward high S/N, but does not impact our source identifications.

In addition to the bias from the steepness of the population, we also treat the fact that the source finder locates the maximum flux along RA and Dec, which provides an extra two degrees of freedom. This can be corrected by finding the flux after two degrees of freedom are subtracted from the detection significance, following \cite{vanderlinde_2010}. Note that this departs from the treatment in \cite{crawford_2010} and \cite{vieira_2010}.  Specifically, all sources are corrected by a factor $(S/N)/\sqrt{S/N^2-2}$.  For sources in the range 20--25\,mJy, this is a correction of 0.5\% at 148\,GHz and 1.5\% at 218\,GHz.  We have confirmed through simulations with synthetic sources implanted in the ACT data that this positional deboosting results in unbiased flux densities.  Furthermore, we have compared raw flux densities from the matched filter to flux densities from the ACT data at the positions of ATCA counterparts, when available.  The latter should not be boosted due to maximizing the flux over position in RA and Dec.  As expected, we find that positional deboosting accounts for the ratio between the raw flux densities and the flux densities derived using ATCA counterpart locations.

As a final consistency check, we note that for the sources observed by both ACT and {\it Planck}, flux densities are consistent at the $\approx$\,1-2\% level, suggesting that the errors assumed here are very conservative and free from any systematic bias (Section \ref{subsec:planckm}).

\subsection{Purity and Completeness}
\label{subsec:purityAndCompleteness}

\hspace{0.2in} The number of false detections at each frequency was estimated by running the detection algorithm on an inverted (negative temperature) map in which we masked the sources and, in the case of the 148\,GHz map (for which the Sunyaev-Zel'dovich effect was non-zero), all ACT-detected and optically confirmed clusters of galaxies.  With this approach, no spurious detections are found in the \arone\ data down to a S/N of 5.  In the \artwo\ data, four spurious detections at S/N\,$\le$\,6 were observed with raw flux densities in the 25--31\,mJy range (Table \ref{tab:stats}).

\begin{table*}
\begin{minipage}{126mm}
\caption{Number Counts, Purity, and Completeness$^{a}$.}
\label{tab:stats}
\begin{tabular}{cccccccccc}
\hline
\hline
                               &  \multicolumn{3}{c}{148 GHz}  &   & \multicolumn{3}{c}{218 GHz} & &   \\   
                  \cline{2-4} \cline{6-8} \\
Flux Range (Jy) & N & Purity  &  Completeness &  &   N & Purity  &  Completeness &   N$_{sync}^{b}$ & N$_{dust}^{c}$  \\
	
\hline
   0.015 -- 0.02 &   23 &  100.0$\pm$0.0\% &    47.1$\pm$3.2\% &  &   8 &  100.0$\pm$0.0\% & 10.7$\pm$2.8\% &  22 &  0 \\ 
    0.02 -- 0.03 &   47 &  100.0$\pm$0.0\% &    75.3$\pm$3.4\% &  &  42 &  92.9$\pm$4.8\% & 34.2$\pm$3.5\% &  46 & 10 \\  
    0.03 -- 0.05 &   48 &  100.0$\pm$0.0\% &    96.8$\pm$1.9\% &  &  43 &  97.7$\pm$2.3\% & 80.2$\pm$3.0\% &  48 & 10 \\  
    0.05 -- 0.09 &   20 &  100.0$\pm$0.0\% &    99.6$\pm$0.6\% &  &  21 &  100.0$\pm$0.0\% & 98.7$\pm$0.9\% &  20 &  1 \\  
    0.09 -- 0.17 &   13 &  100.0$\pm$0.0\% &   100.0$\pm$0.0\% &  &  12 &  100.0$\pm$0.0\% & 100.0$\pm$0.0\% &  13 &  2 \\  
    0.17 -- 0.33 &    4 &  100.0$\pm$0.0\% &   100.0$\pm$0.0\% &  &   2 &  100.0$\pm$0.0\% & 100.0$\pm$0.0\% &   4 &  0 \\   
    0.33 -- 0.65 &    2 &  100.0$\pm$0.0\% &   100.0$\pm$0.0\% &  &   1 &  100.0$\pm$0.0\% & 100.0$\pm$0.0\% &   2 &  0 \\  
    0.65 -- 1.39 &    1 &  100.0$\pm$0.0\% &   100.0$\pm$0.0\% &  &   1 &  100.0$\pm$0.0\% & 100.0$\pm$0.0\% &   1 &  0 \\  
    1.39 -- 2.87 &    1 &  100.0$\pm$0.0\% &   100.0$\pm$0.0\% &  &   1 &  100.0$\pm$0.0\% & 100.0$\pm$0.0\% &   1 &  0 \\  
\hline
\end{tabular}
\medskip
$^{a}$The number of sources at each frequency with S/N\,=\,5 over 455\,deg$^2$.  Errors on number count are simply Poisson.  See Figure \ref{fig:dNdS} for a graph of purity/completeness-corrected differential source counts.\\
$^{b}$Counts of synchrotron-dominated sources are taken relative to the 148\,GHz flux density.\\
$^{c}$Counts of dust-dominated sources are binned according to their 218\,GHz flux density.
\end{minipage}
\end{table*}

The purity simulations appear to be consistent with our findings from source cross-identification (Section \ref{subsec:others}).  At 148\,GHz, only one 5$\sigma$ source, ACT-S J023600-530237, is not cross-identified with other catalogs.  At 218\,GHz, five 5$\sigma$ sources, ACT-S J024430-541605, ACT-S J035034-524801, ACT-S J062747-512614, ACT-S J063715-500414 and ACT-S J065207-551605, are not cross-identified with other catalogs, although we expect some of these to be real DSFGs, a hypothesis which will be tested through future follow-up observations.

In order to estimate completeness of the catalog for each band, we added 100 synthetic sources of a single flux density to the 455\,deg$^2$ region of the ACT maps used in this study. The map filtering and source detection algorithm (Section \ref{sec:sourceExtraction}) were run, and the resulting catalog checked for inclusion of the input sources. This procedure was repeated in each band in intervals of 10\,mJy in synthetic source flux density in the range 10--100\,mJy.  The completeness at each flux density was interpolated to estimate the completeness in flux density bins given in Table \ref{tab:stats}.

\section{Comparison to Other Catalogs}
\label{subsec:others}

\begin{figure}
  \begin{center}
  \includegraphics[scale=0.69, clip, trim=8.5mm 7mm 6mm 6mm]{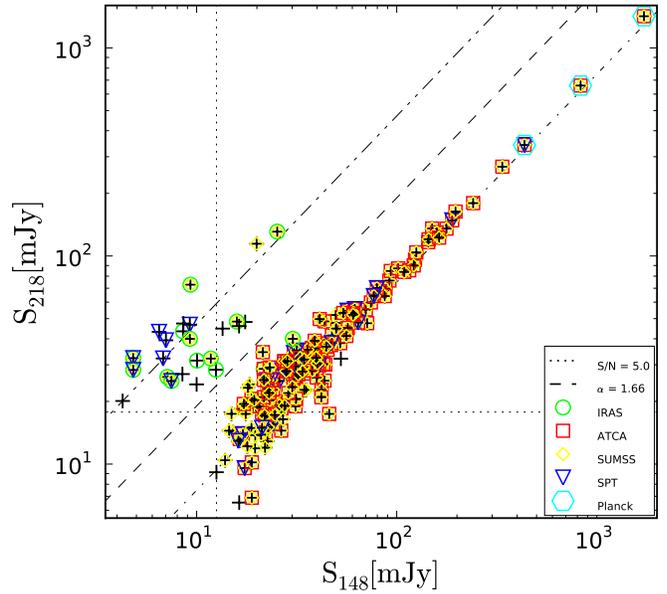}
  \end{center}
  \caption[The ACT sources cross-identified in flux density space.]{ The ACT sources cross-identified in flux density space.  Sources have been selected with S/N\,$>$\,5.0 in at least one frequency band (dotted lines).  See text for a more thorough description of cross-identification statistics with other catalogs.  The diagonal lines follow the power law $S_{\nu}\,\propto\,\nu^{\alpha}$ with $\alpha$\,=\,1.66 for the dashed line, distinguishing the $\alpha$\,=\,$-0.6$ (dash-dot) synchrotron-dominated source population from the $\alpha$\,=\,3.7 (dash-dot-dot) dust-dominated source population.}
  \label{fig:crossIDs}
\end{figure}

\hspace{0.2in} To further characterize ACT sources, we consider cross-identifications with other catalogs.  Given the number of ACT-detected sources and the ACT beam size, around 1 out of 10,000 sources randomly placed in the 455\,deg$^2$ area considered here would coincide with an ACT source.  Figure \ref{fig:crossIDs} shows the ACT sources located in flux density space, and cross-identifications from some overlapping catalogs with extragalactic source detections in the southern hemisphere discussed in this section.  The synchrotron-dominated sources occupy points near the power law $S_{\nu}\,\propto\,\nu^{\alpha}$ with spectral index $\alpha$\,=\,-0.6.  Dust-dominated sources will follow a line closer to $\alpha$\,=\,3.7, and are generally associated with sources detected with S/N\,$>$\,5 at 218\,GHz only.

As our cross identifications come from radio catalogs at low frequencies, and dust-dominated sources lack strong synchrotron radiation, we expect fewer of the dust-dominated sources to be co-identified with a previously known radio source.  Overlapping large-area, sensitive far IR catalogs would be the equivalent for our dust-dominated sources, but these do not exist.  Even if they did, they would be too confused for positive cross-identification in this manner.  At present, these unidentified sources are the subject of individual follow-up studies \citep[e.g.,][]{ikarashi_2010, scott_2011, lupu_2012, weil_2013}.

\subsection{General Statistics of Identifications}

\hspace{0.2in} Of our \numberOfSources\ sources, many have cross-identifications with several catalogs.  One hundred seventy-four are identified with sources in the 0.84\,GHz Sydney University Molonglo Sky Survey \citep[SUMSS;][]{mauch_2003}, and 122 of those cross-identified sources also belong to the 4.85\,GHz Parkes-MIT-NRAO (PMN) survey radio catalog \citep{wright_1994}.  Fourteen sources were cross-identified with the Infrared Astronomical Satellite \citep[IRAS;][]{helou_1988} at 12--100\,$\mu$m.   Sources were also identified with one or both of the Australian Telescope 20\,GHz survey \citep[AT20G;][]{murphy_2010} and the 1.4/2.0\,mm SPT \citep{vieira_2010} catalogs.  Two of the remaining 8 unmatched sources were observed during the November 2010 ATCA follow-up campaign (Section 2.3).  This leaves 6 sources (ACT-S J023600-530237, ACT-S J024430-541605, ACT-S J035034-524801, ACT-S J062747-512614, ACT-S J063715-500414 and ACT-S J065207-551605) with no identification in any of the available catalogs, all but one of which have dust-dominated spectra.  The single unidentified source with a synchrotron spectrum, ACT-S J023600-530237, falls near our S/N\,=\,5 threshold at 148\,GHz and could be a false detection.

A NASA/IPAC Extragalactic Database (NED) search using a radius of 2\arcmin\ cross-identified 3 sources from the ACT catalog with galaxy clusters: Abell S0250, Abell 3391 and S\'{e}rsic 037/01.  Increasing the search radius to 3\arcmin\ adds Abell 3128 and Abell 3395 to this list.  Simulations of the microwave sky suggest that only $\approx$\,3\% of galaxy clusters have their 148\,GHz signal contaminated at 20\% or more \citep{sehgal_2010}. 

\subsection{Matches to AT20G}

\hspace{0.2in} The AT20G survey, carried out from 2004 to 2008 at 20\,GHz with follow-up at 5 and 8\,GHz, covered 6.1\,sr of the Southern sky to a flux density limit of 40\,mJy at 20\,GHz \citep{murphy_2010}.  Of our sources, 115 match sources in the AT20G catalog, listed in Table~4, all of which we classify as synchrotron-dominated.   Given the AT20G completeness limit and the predominance of flat spectra for our 148\,GHz-selected sources (Section \ref{subsec:spectra}), faint ACT sources may not have matches in AT20G.  We received time on the Australia Telescope Compact Array to measure flux densities for sources in the earlier M11 catalog that did not appear in AT20G, the results of which are described in Sections \ref{subsec:atcaobs} and \ref{subsec:spectra}.

\subsection{Matches to the Planck Compact Source Catalogs}
\label{subsec:planckm}

\hspace{0.2in} The {\it Planck} satellite team has released two sets of all-sky catalogs in nine frequency channels, including two very close to those used by ACT.  The first set of catalogs made up the Early Release Compact Source Catalog \citep[ERCSC;][]{planck13_2011}.  Very recently, a deeper catalog based on 1.6 complete {\it Planck} surveys has been released, the Planck Catalog of Compact Sources \citep[PCCS;][]{planck28_2013}.  At 143 and 217\,GHz, the {\it Planck} beam sizes are approximately 7\arcmin\ and 5\arcmin\ respectively.   Like WMAP, {\it Planck} makes great gains by observing the entire sky simultaneously at several frequencies, thus capturing all bright sources.  On the other hand, due to their large beam sizes, these experiments detect {\it only} the brightest sources.   In the case of the PCCS, the detection threshold at 143 and 217\,GHz is $\approx$\,400\,mJy; at lower flux densities the catalog completeness drops well below 80\%.   In the area treated in this study, {\it Planck} detects 3 of the brightest ACT sources. 

For a more complete comparison with the PCCS, we examined all bright ACT sources, combining the Equatorial and Southern regions mapped by ACT.  Almost 50 matches were found in this wider comparison.  At 148\,GHz, the agreement of ACT flux densities and the PCCS flux densities, properly color-corrected and (slightly) extrapolated to match ACT's central frequency, is excellent, at the 1-2\% level, despite scatter introduced by source variability.  At 218\,GHz, source variability plays at least as large a role, and PCCS flux densities are $\approx$\,5\% higher than ACT's.  If we remove a couple of the most variable sources, however, the agreement of the 218\,GHz flux density scales improves to $\approx$\,1\%.  At both frequencies, {\it Planck} fluxes are on average slightly higher.  Agreement at the $\approx$\,1-2\% level, however, is well within the expected uncertainty in the ACT flux densities (Section \ref{sec:data}).  This agreement suggests that our flux density values are free of systematic error.

\subsection{Matches to the SPT Catalog}

\hspace{0.2in} The SPT study of compact sources \citep{vieira_2010} was based on observations of a square patch of sky of 87\,deg$^2$ centered at 05$^h$ right ascension, having only fractional ACT overlap. Nevertheless, 2,304 of the 3,496 SPT candidate sources (those with S/N\,$>$\,3 and flux densities $>$\,4.4\,mJy) fall within the ACT survey region.  We find 32 cross-identifications with ACT sources when we search the SPT 1.4\,mm (220\,GHz) and 2.0\,mm (145\,GHz) catalogs.  Twenty-five of these were 148\,GHz-selected and categorized in \cite{vieira_2010} as synchrotron-dominated.  The remaining seven were detected by ACT with S/N\,$>$\,5 only at 218\,GHz; three of the seven also have SUMSS or IRAS cross-identifications.  To make a direct comparison of the flux densities of ACT-SPT cross-identified sources, we used our derived flux densities, but without introducing two angular degrees of freedom in the deboosting \citep[as in][]{vanderlinde_2010}, as these were not incorporated by the SPT analysis \citep{crawford_2010}.  We find an overall offset at the 7\,$\pm$\,5\% level at 148\,GHz and 3\,$\pm$\,7\% at 218\,GHz, with ACT flux densities typically higher than those of SPT.  At low flux, the ACT and SPT measurements of individual sources agree within their errors. However, at fluxes above 30\,mJy, individual sources may disagree at several sigma. We understand the discrepancy to be due to variability in radio sources, which is deferred to a future publication.

Recently, using spectroscopic follow-up observations with the Atacama Large Millimeter/submillimeter Array (ALMA), \cite{vieira_2013} and \cite{weil_2013}, derived robust redshifts for about 18 of the SPT-detected dust-dominated sources from a larger sky area than in \cite{vieira_2010}.  One of these sources, found to be at redshift 5.66, matches ACT-S J034640-520505 which otherwise had no cross-identification.  A second previously unmatched source, ACT-S J002707-500713, was imaged with ALMA at 870\,$\mu$m, is resolved at the 0.5\arcsec\ scale, likely due to gravitational lensing, but awaits spectroscopic follow-up.  Four other ACT dust-dominated sources that were cross-identified with the \cite{vieira_2010} catalog now also have robust redshifts.  ACT-S J055139-505800 is at redshift 2.123, ACT-S J053250-504709 is at redshift 3.399, ACT-S J052903-543650 is a source at redshift 3.369, with a lens at redshift 0.140, and ACT-S J053817-503058 has a redshift of 2.782, and a lens at a redshift of 0.404.  It is likely that the few ACT dust-dominated sources which lie outside the SPT footprint are also lensed DSFGs.

\subsection{Matches to the IRAS Catalog}

\hspace{0.2in} Given the ACT beam size, a normal galaxy will be unresolved at redshifts $z$\,$\ge$\,0.05 or distances greater than 200\,Mpc.  Consequently, only very nearby objects appear extended in our maps.  The IRAS source population consists primarily of local ($z$\,$\ll$\,1) dust-dominated ULIRGs; we thus expect our resolved sources to coincide with IRAS sources.  We find 14 sources cross-identified with IRAS sources, out of a possible 829 IRAS sources that lie within the ACT survey area treated here.  Two of these have synchrotron-dominated spectra, both of which are nearby galaxies.  The twelve that show spectra dominated by dust re-emission are Galactic sources, in the Magellanic clouds, or are known nearby star-forming galaxies.  For example, for two local resolved galaxies particularly bright at ACT frequencies, NGC1566 (ACT-S J041959-545622), a Seyfert two-arm spiral, and IC1954 (ACT-S J033133-515352), a late-type spiral with a short central bar, ACT observes a higher 218\,GHz flux density than at 148\,GHz or 20\,GHz, confirming the dust-dominated nature of their spectra.

\subsection{Dust-Dominated ACT Sources}

\hspace{0.2in} Of the 24 ACT dust-dominated sources in the catalog presented here, 18 of these are cross-identified with either IRAS (12 sources) and/or SPT (9 sources, including the recent ALMA observations).  ACT-S J051506-534420, ACT-S J053311-523827, and ACT-S J055115-533435 were cross-identified with both catalogs.  Ten of the dust-dominated sources are cross-identified with sources in the SUMSS catalog, bringing the total for cross-identified sources to 19.  The remaining 5 sources, ACT-S J024430-541605, ACT-S J035034-524801, ACT-S J062747-512614, ACT-S J063715-500414 and ACT-S J065207-551605, have no matching counterpart, and signal-to-noise at 218\,GHz in the range 5.27\,$<$\,S/N\,$<$\,6.35.  Purity tests suggest that a couple of these could be spurious.  However, given that three S/N\,$\approx$\,6 sources are cross-identified with SPT, some of these sources are likely real detections, high redshift galaxies lensed by intervening structure.  Future follow-up observations are planned to clarify the nature of these detections.

The main and supplementary 500\,$\mu$m samples from the {\it Herschel} HerMES survey have lensing candidate densities of 0.14\,$\pm$\,0.04\,deg$^{-2}$ and 0.31\,$\pm$\,0.06\,deg$^{-2}$, respectively \citep{wardlow_2013}.  An analogous sample from {\it Herschel}'s H-ATLAS survey is 0.35\,$\pm$\,0.16\,deg$^{-2}$ \citep{negrello_2010}. In the same bands as presented here, SPT finds 0.25\,$\pm$\,0.02\,deg$^{-2}$ lensing candidates \citep{mocanu_2013}. This suggests that $\approx$\,100 such sources exist in the field here, of which we see the high flux tail.

\section{Source Counts}
\label{subsec:counts}

\begin{figure*}
  \begin{center}
  \includegraphics[scale=0.98, clip, trim=3mm 0mm 4mm 4mm]{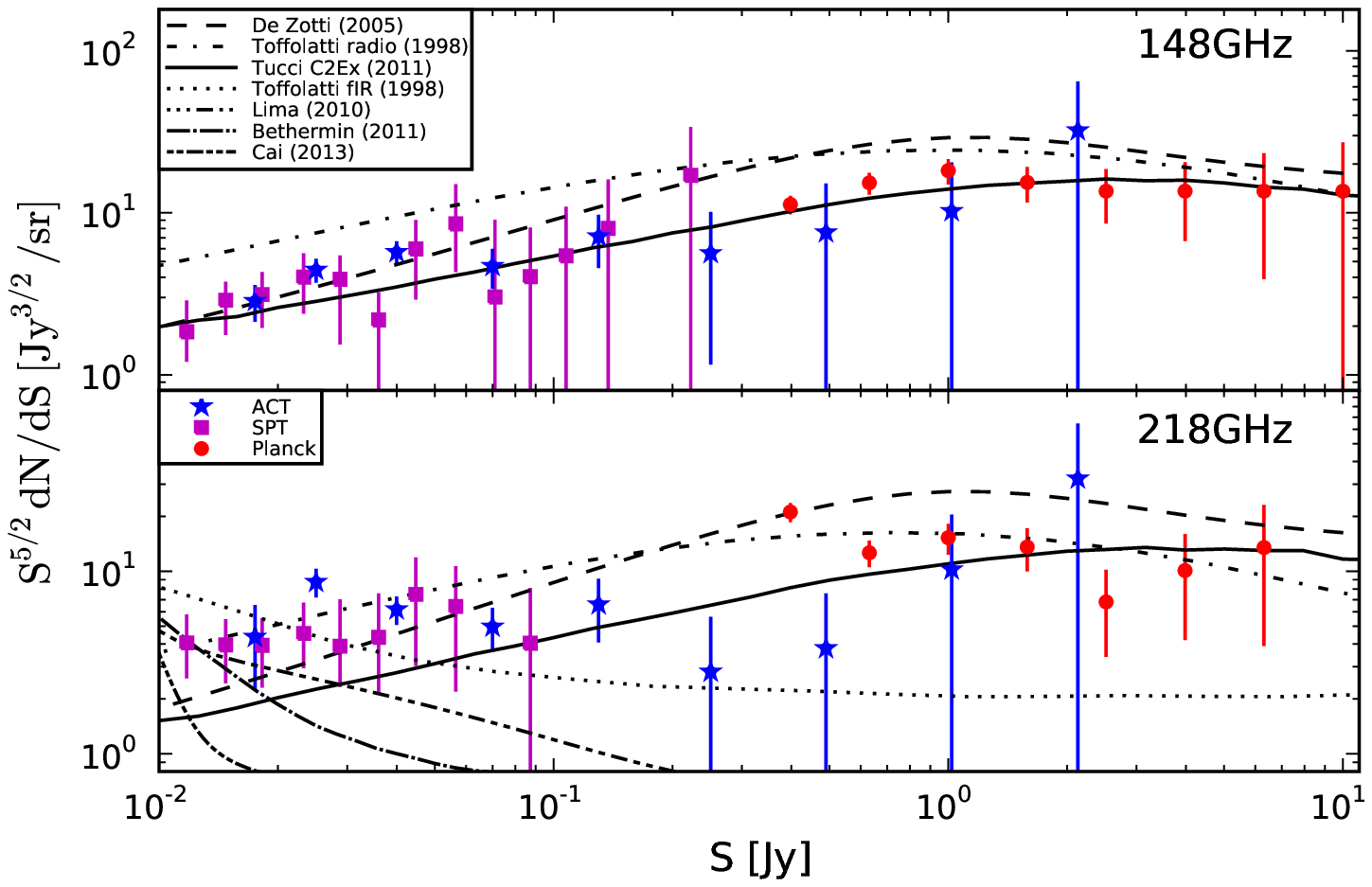}
    \end{center}
  \caption[Differential number counts of ACT-selected sources.]{ Differential number counts of ACT-selected sources.   Derived from Table \ref{tab:stats} and corrected for completeness, the ACT differential source counts are plotted together with models of radio and infrared source populations.   The \cite{planck7_2012} data points and the SPT data points of \cite{vieira_2010} are also plotted, both of which are consistent with ACT counts.  The ACT and Planck points have Poissonian errors (1$\sigma$), whereas the SPT points include measurement and independent calibration errors.  The fIR models on this scale predicts number counts too low to be seen at 148\,GHz. The ACT data are consistent with being dominated by radio sources at both frequencies.
}
  \label{fig:dNdS}
\end{figure*}

\hspace{0.2in} The completeness-corrected differential number counts for ACT sources based on the data in Table \ref{tab:stats} are plotted in Figure \ref{fig:dNdS}.  For comparison, completeness-corrected number counts from \cite{vieira_2010} and \cite{planck13_2011} are plotted as well.  The ACT data fill in the flux density gap between the SPT and {\it Planck} catalogs at these frequencies, caused by the differences between experiments in sky coverage and sensitivity to point sources. The effect of calibration error is to shift the flux bins by $\pm$\,2\% ($\pm$\,2.4\%) at 148\,GHz (218\,GHz; Section \ref{sec:data}).

The combined ACT, SPT and {\it Planck} total counts at 148\,GHz and synchrotron-dominated source counts at 218\,GHz are best fit by the recently developed \cite{tucci_2011} C2Ex radio source model. This model divides the $\sim$\,flat- or steep-spectrum blazar population into BL Lac objects, with a region close to the AGN core dominating the observed emission ($\leq$\,0.3\,pc), and flat spectrum radio quasars (FSRQs) with a break frequency indicating that the source of emission arises from an emitting region further from the AGN core (0.3 to 10\,pc).  The \cite{dezotti_2005} model for radio source counts, while consistent with the counts below 0.1\,Jy, over-predicts the counts at higher flux densities.   {\it Planck} analysis finds that the \cite{dezotti_2005} model is consistent with their counts at frequencies up to 100\,GHz, but over-predicts the counts at higher frequencies in the flux density region of $\approx$\,1\,Jy, though they begin to suffer from incompleteness below $\approx$\,1\,Jy. Our few $\approx$\,1\,Jy brightest sources appear to be consistent with this finding, but lend little statistical significance.

As well as the models for counts of radio sources, the source counts predicted for dusty starburst galaxies from \cite{toffolatti_1998},  \cite{lima_2010}, \cite{bethermin_2011} and \cite{cai_2013} are shown in Figure \ref{fig:dNdS}.  The brightest infrared sources in the \cite{toffolatti_1998} model are 10\,mJy. Given that all sources in the ACT catalog have flux densities greater than 10\,mJy, these models predict that the 148\,GHz- and 218\,GHz-selected ACT catalogs should have few or no detected infrared sources.  However, we find 24 sources that have dust-dominated spectra with flux densities above 20\,mJy, and \cite{vieira_2010} find 36 dust-dominated sources with flux densities above 15\,mJy.  While \cite{toffolatti_1998} does not incorporate the effects of lensing on observed number counts, the models of \cite{lima_2010}, \cite{bethermin_2011}, and \cite{cai_2013} modelled dust-dominated galaxies together with lensed high-redshift dust-dominated galaxies.  However, the small number of dust-dominated sources present in Figure \ref{fig:dNdS} lend little constraining power, and we leave it to future larger-area ACT studies to analyze the robustness of these models.  The \cite{bethermin_2011} model, falling roughly in the middle of this group of models, was used for our flux deboosting.  

\section{Source Spectra}
\label{subsec:spectra}

\begin{table*}
\begin{centering}
\begin{minipage}{126mm}
\caption{Median Spectral Indices.}
\label{tab:spectra}
\begin{tabular}{ccccc}
\hline
\hline
Spectral Index & Synchrotron (all)$^{a}$ & Synchrotron (S$_{148}$\,$>$\,50\,mJy) & Synchrotron (S$_{148}$\,$<$\,50\,mJy) & Dust-dominated\\
\hline
$\alpha_{5-20}$	    & 	-0.15$^{+0.37}_{-0.36}$ &   -0.07$^{+0.36}_{-0.25}$   &    -0.21$^{+0.33}_{-0.41}$   &  ...  \\ 
$\alpha_{20-148}$       & 	-0.42$^{+0.32}_{-0.26}$ &   -0.36$^{+0.24}_{-0.29}$   &    -0.43$^{+0.34}_{-0.33}$   &  ...   \\
$\alpha_{148-218}$     & 	-0.55$^{+0.60}_{-0.60}$ &   -0.60$^{+0.20}_{-0.20}$   &    -0.51$^{+0.38}_{-0.71}$   &  3.7$^{+0.62}_{-0.86}$   \\
\hline
\end{tabular}
\medskip
$^{a}$Quoted errors are the 68\% confidence levels of the distribution.
\end{minipage}
\end{centering}
\end{table*}

\hspace{0.2in} Source spectral energy distributions (SEDs) can be used to differentiate source types by their dominant emission mechanisms.  Assuming the commonly used simple power law model $S(\nu)\,\propto\,\nu^{\alpha}$, a negative $\alpha$ is indicative of sources dominated by synchrotron emission, such as radio galaxies.  Sources with free-free emission dominating will have an index close to 0.  The high-redshift SMG population will have spectra dominated by re-emission of their prodigious optical and UV flux by the surrounding dust in a greybody spectrum, with indices expected to be greater than 2 and more typically 3-4.  

We can divide the ACT source population according to several broad spectral groups: classical steep (and steepening) spectrum sources, sources that peak within the frequency range under consideration, and sources that show flat, rising or upturned spectra.  

The ACT catalog is dominated by synchrotron-dominated blazers, which have variable flux densities.  This variation in flux density is not periodic.  For any single source, then, inferences about its spectrum will depend on the epoch of observation, although not biased one way or the other.  For the catalog as an ensemble, however, a spectral study may give rise to insights about the average spectral behavior of the galaxy populations. 

Table \ref{tab:spectra} summarizes the median spectral indices between pairs of frequencies for various subsets of the ACT data.  In obtaining the spectral indices we compare the deboosted flux densities from ACT with the raw flux densities from AT20G.  The whole sample includes data from the 20\,GHz ATCA follow-up observations taken in November 2010 (Section \ref{subsec:atcaobs}).  

\begin{figure}
\begin{center}
	\includegraphics[scale=.47, clip, trim=6mm 11mm 2mm 20mm]{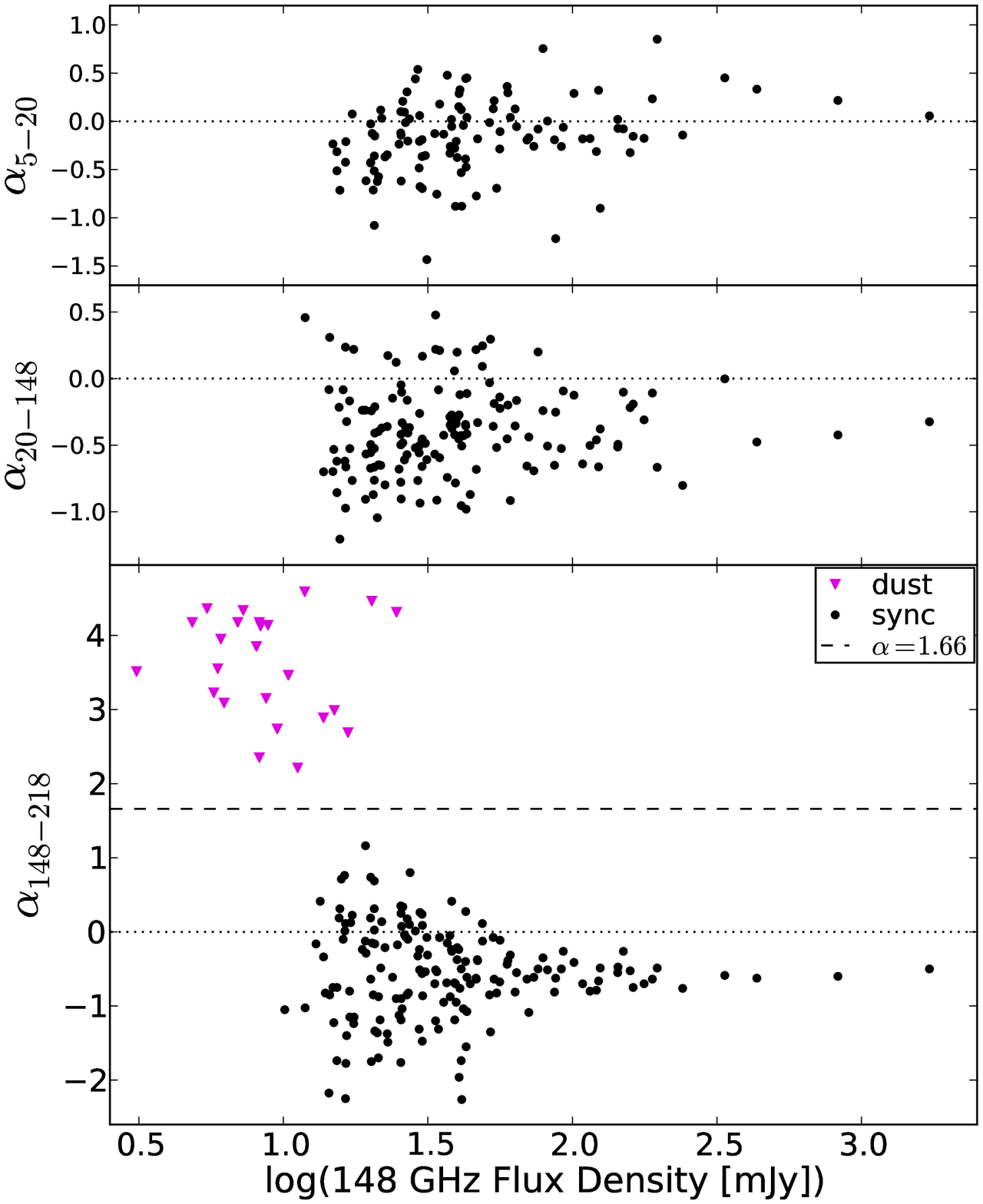}
  \end{center}
 \caption{Color-magnitude diagrams comparing the 5--20\,GHz (top), 20--\arone\ (middle), and 148--\artwo\ (bottom) spectral indices for ACT-ATCA cross-identified sources. The synchrotron-dominated radio galaxy population is dominated by sources which have consistently falling SEDs towards higher frequencies. }
\label{fig:spectralIndices}
\end{figure}

Figure \ref{fig:spectralIndices} shows the 5--20\,GHz, 20--148\,GHz and 148--218\,GHz color-magnitude diagrams. The radio-selected AGN (black points) are predominantly characterized by steepening of the spectra, with only a few characterized by extremely flat or inverted spectra.  There is a clear trend towards more negative median spectral index with increasing frequency.  Magenta points, showing rising spectra between 148 and 218\,GHz, denote sources ACT-classified as spectrally dust dominated.  They do not show up in the top two plots, indicating that these sources have flux densities falling below the detection threshold of the AT20G catalog.   

The average spectral indices between AT20G and ACT frequencies indicates an underlying source population made up of Flat-Spectrum Radio Quasars (FSRQ), a type of blazar, with AGN jet pointed along our line of sight \citep{dezotti_2010}.  Ejected material flows through several shocked regions in the jet which locally enhance the radiation \citep{marscher_1985, valtaoja_1992}. The observed spectral flatness is the superposition of many components with different turnover frequencies.  At frequencies greater than approximately 100\,GHz, however, \cite{marriage_2011a}, \cite{vieira_2010}, and \cite{planck13_2011} observe a steepening of the spectrum ($\alpha_{148-218}$\,$\approx$\,-0.6) that until now had not been conclusively shown \citep{tucci_2011}.  This change is possibly due to electron energy losses in the jet (``electron ageing'') or the transition to the optically thin regime in the extended radio lobes.  The underlying physical mechanisms have been contested for more than a decade and remain the subject of ongoing study,  \citep[e.g.,][]{nieppola_2008, ghisellini_2008, sambruna_2010}.

\begin{figure}
\begin{center}
    \includegraphics[scale=0.63, clip, trim=2mm 1mm 5mm 5mm]{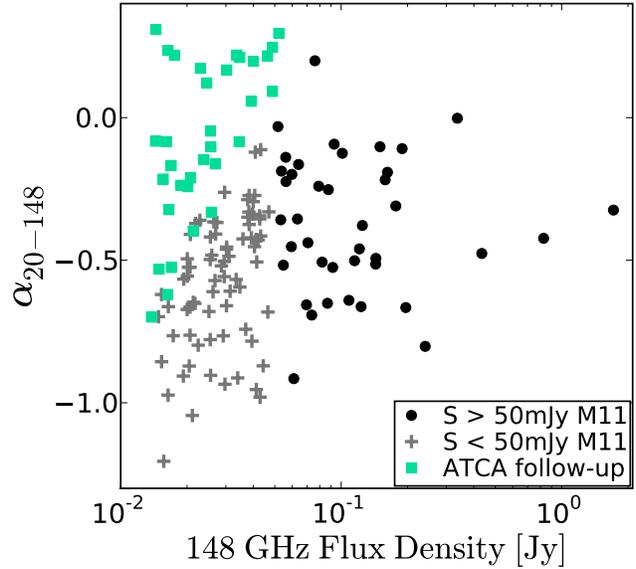}
 \caption{ Radio color-magnitude diagram using 20--148\,GHz spectral indices for ACT-AT20G cross-identified sources. The data are divided between flux densities at 50\,mJy at 148\,GHz. The low flux sample was incomplete and suffered from selection bias that favored sources with more negative spectral indices. Data from the ATCA follow-up study for the M11 ACT 148\,GHz sources not cross-identified with AT20G are shown as green squares. The high flux sample, denoted with black points, has a median spectral index -0.37. Prior to ATCA follow-up observations, the lower flux ACT sources cross-identified with the AT20G catalog (grey crosses) had a median spectral index of -0.52.  Including the follow-up data, the more complete lower flux sample has a median spectral index of -0.43. }
\label{fig:color-magNewdata}
\end{center}
\end{figure}

All but one of the synchrotron-dominated ACT detections with flux density $>$\,50\,mJy have cross-identifications in AT20G.  Below this flux density, the mean spectral indices of the population of ACT-AT20G cross-identified sources are biased toward the negative by the incompleteness of the AT20G catalog below 100\,mJy at 5\,GHz and below 40\,mJy at 20\,GHz.  It is therefore illustrative to further divide this subpopulation according to 148\,GHz flux density.  

Figure \ref{fig:color-magNewdata}, a radio color-magnitude diagram, plots spectral index $\alpha_{20-148}$ against 148\,GHz source flux density.  Black and grey points identify the M11 sources that were cross-identified with the AT20G catalog.  Black points are for sources with flux densities $>$\,50\,mJy, which represented a complete sample.  Grey points denote the fainter, incomplete sample.  The blue points were obtained by calculating the spectral index for the previously unmatched, lower flux density subsample, followed up with ATCA (see Table~4 for AT2G0 source IDs with an asterisk).  The population represented by the blue points fills in the picture remarkably for the fainter (below 50mJy flux density at 148\,GHz, below 40\,mJy at 20\,GHz) population, especially in the region with spectral index of approximately zero. 

The unbiased $S_{148}$\,$>$\,50\,mJy sample has a 20--148\,GHz median spectral index of -0.36$^{+0.24}_{-0.29}$.  For the $S_{148}$\,$<$\,50\,mJy subsample prior to follow-up with ATCA (grey points only), the 20--148\,GHz spectral index was $\alpha_{20-148}$\,=\,-0.52$^{+0.17}_{-0.28}$.  However, with the addition of the synchrotron-classified ACT-selected sources with 2010 ATCA follow-up data (grey and blue points), the full sample has 20--148\,GHz spectral index $\alpha_{20-148}$\,=\,-0.42$^{+0.32}_{-0.26}$.  This supports the hypothesis that the lower flux contingent is probing the same population of synchrotron-dominated sources (blazers), and that there isn't much evolution of their spectral index with flux.  The full dataset suggests that the increased scatter at lower fluxes is due to variability (which will have a larger relative effect on the fluxes) and decreased S/N from flux error.

For our dust-dominated sources, we find a median spectral index of $\alpha_{\mathrm{148-218}}$\,$=$\,3.7$^{+0.62}_{-0.86}$.  For sources detected above 5$\sigma$ at both 150 and 220\,GHz, SPT derives $\alpha_{\mathrm{150-220}}$\,$=$\,3.35\,$\pm$\,0.7 \citep{mocanu_2013}.  These indices are consistent with the findings of \cite{dunkley_2011}, where the effective index of unresolved IR emission as determined from ACT data is 3.69\,$\pm$\,0.14, and the best-fit mean spectral indices of $\alpha^P_{\mathrm{150-220}}$\,$=$\,3.86\,$\pm$\,0.23 for Poisson-distributed sources, and $\alpha^C_{\mathrm{150-220}}$\,$=$\,3.80\,$\pm$\,1.3 for the clustered component, found below the SPT detection threshold \citep{hall_2010}, a population expected to be dominated by dust-dominated sources.  If we model dust emission as a modified blackbody, in the Rayleigh-Jeans (RJ) limit the dust emissivity index is related to the effective spectral index as $\beta$\,$=$\,$\alpha$\,$-$\,2\,$=$\,1.7, consistent with models of $\approx$\,30K dust made from graphite and silicate grains \citep{draine_1984}.  We leave a more rigorous analysis involving redshifted greybodies, where the RJ limit is not as good an approximation, and the implications for star formation, to future work.


\section{Conclusion}
\label{sec:conc}

\hspace{0.2in} We have described the extragalactic source population at 148 and 218\,GHz found in a 455\,deg$^2$ region of the ACT 2008 \emph{Southern strip}, centered on declination -52.5$\degr$. This updates the ACT 148\,GHz catalog by using the data released with \cite{dunner_2013}, extends the results in \cite{marriage_2011a} to two bands, and treats noise as more local, which in turn yields a higher S/N. We detect 191 sources above S/N\,=\,5 in at least one of the ACT 148 and 218\,GHz frequency bands, spanning flux densities 14--1700\,mJy (Table~4).  Known redshifts are as high as $\approx$\,6, with measurements ongoing. The catalog is estimated to be 100\% pure and 96.8\% complete above 30\,mJy at 148\,GHz, and 97.7\% pure and 80.2\% complete above 30\,mJy at 218\,GHz. We have confirmed flux recovery of the pipeline, and jointly deboosted the flux densities in both bands.

The multifrequency nature of our observations allows for internal classification of sources into three broad classes of sources based on their spectra: synchrotron-dominated sources (the vast majority of which are cross-identified with radio catalogs), low-redshift dust-dominated sources with IRAS counterparts (typically ULIRGs), and dust-dominated sources also observed by SPT or with no cross-identification, either shown or assumed to be high redshift star-forming galaxies. This last class, only recently observed at millimeter wavelengths, has many of the properties expected of the progenitor population of massive, modern-day, elliptical galaxies, background SMGs whose flux has been magnified through gravitational lensing by a foreground galaxy or galaxy group.  This interpretation, bolstered by population synthesis analyses \citep[e.g.][]{thomas_2005}, is being validated with follow-up observations. 

A comparison with other catalogs shows that 97\% of ACT-detected sources correspond to sources detected at lower or higher frequencies. The 148\,GHz source counts are fit reasonably well by the C2Ex radio model of \cite{tucci_2011}, the most current model for radio sources.  According to the analysis of the average spectral indices derived from the combined AT20G and ACT datasets, the ACT data support the case for a spectral steepening toward higher frequencies above 100\,GHz for AGN. The ACT dust-dominated source population has a median spectral index, $\alpha_{\mathrm{148-218}}$, of 3.7$^{+0.62}_{-0.86}$.  Properly linking these sources into the broader context of galaxy formation and evolution is of cosmological interest, and a goal of future work. 

The analysis presented here uses only the 2008 data of ACT's \emph{Southern strip}, representing only 1/6$^{th}$ of the data ACT obtained between 2007 and 2010. In future work, we will extend our analysis of the source population to include the full dataset integrating both the southern and equatorial regions observed by ACT.  The ACT \emph{Equatorial strip} overlaps with deep Sloan Digital Sky Survey Stripe 82 observations \citep{annis_2011}.  Thus as well as increasing the sky coverage and number counts for the ACT sources, future work (Gralla et al. in prep.) will include joint analyses with optical data.

\section*{Acknowledgments}

\hspace{0.2in} This work was supported by the U.S. National Science Foundation through awards AST-0408698 and AST-0965625 for the ACT project, and PHY-0355328, PHY-0855887, PHY-1214379, AST-0707731 and PIRE-0507768 (award number OISE-0530095).  Funding was also provided by Princeton University, the University of Pennsylvania, and a Canada Foundation for Innovation (CFI) award to UBC.  ES acknowledges support by NSF Physics Frontier Center grant PHY-0114422 to the Kavli Institute of Cosmological Physics.  The PIRE program enabled this research program through exchanges between Chile, South Africa, Spain and the US.  Computations were performed on the GPC supercomputer at the SciNet HPC Consortium.  SciNet is funded by: SciNet is funded by the Canada Foundation for Innovation (CFI) under the auspices of Compute Canada, the Government of Ontario, the Ontario Research Fund -- Research Excellence, and the University of Toronto. Data acquisition electronics were developed with assistance from the CFI.  ACT operates in the Parque Astron—mico Atacama in northern Chile under the auspices of the Comisi\'on Nacional de Investigaci\'on Cient\'ifica y Tecnol\'ogica de Chile (CONICYT).

We thank the staff at the Australia Telescope Compact Array site, Narrabri (NSW), for the valuable support they provide in running the telescope. ATCA is part of the Australia Telescope National Facility which is funded by the Commonwealth of Australia for operation as a National Facility managed by CSIRO.  

This research made use of the NASA/IPAC Extragalactic Database (NED) which is operated by the Jet Propulsion Laboratory, California Institute of Technology, under contract with the National Aeronautics and Space Administration.

ACT data products are publicly accessible through LAMBDA (http://lambda.gsfc.nasa.gov/) and the ACT website (http://www.physics.princeton.edu/act/).

\bibliographystyle{mn2e}
\bibliography{act}

\begin{thebibliography}{}

\bibitem[\protect\citeauthoryear{{Amblard}, {Cooray}, {Serra}, {Temi},
  {Barton}, {Negrello}, {Auld}, {Baes}, {Baldry}, {Bamford}, {Blain}, {Bock},
  {Bonfield}, {Burgarella}, {Buttiglione}, {Cameron}, {Cava}, {Clements} \&
  {Croom}}{{Amblard} et~al.}{2010}]{amblard_2010}
{Amblard} A.,  {Cooray} A.,  {Serra} P.,  {Temi} P.,  {Barton} E.,  {Negrello}
  M.,  {Auld} R.,  {Baes} M.,  {Baldry} I.~K.,  {Bamford} S.,  {Blain} A.,
  {Bock} J.,  {Bonfield} D.,  {Burgarella} D.,  {Buttiglione} S.,  {Cameron}
  E.,  {Cava} A.,  {Clements} D.,    {Croom} S. e.~a.,  2010, A\&A, 518, L9+

\bibitem[\protect\citeauthoryear{{Annis}, {Soares-Santos}, {Strauss}, {Becker},
  {Dodelson}, {Fan}, {Gunn}, {Hao}, {Ivezic}, {Jester}, {Jiang}, {Johnston},
  {Kubo}, {Lampeitl}, {Lin}, {Lupton}, {Miknaitis}, {Seo}, {Simet} \&
  {Yanny}}{{Annis} et~al.}{2011}]{annis_2011}
{Annis} J.,  {Soares-Santos} M.,  {Strauss} M.~A.,  {Becker} A.~C.,  {Dodelson}
  S.,  {Fan} X.,  {Gunn} J.~E.,  {Hao} J.,  {Ivezic} Z.,  {Jester} S.,  {Jiang}
  L.,  {Johnston} D.~E.,  {Kubo} J.~M.,  {Lampeitl} H.,  {Lin} H.,  {Lupton}
  R.~H.,  {Miknaitis} G.,  {Seo} H.-J.,  {Simet} M.,    {Yanny} B.,  2011,
  ArXiv e-prints

\bibitem[\protect\citeauthoryear{{Austermann}, {Dunlop}, {Perera}, {Scott},
  {Wilson}, {Aretxaga}, {Hughes}, {Almaini}, {Chapin}, {Chapman}, {Cirasuolo},
  {Clements}, {Coppin}, {Dunne}, {Dye}, {Eales}, {Egami} \&
  {Farrah}}{{Austermann} et~al.}{2010}]{austermann_2010}
{Austermann} J.~E.,  {Dunlop} J.~S.,  {Perera} T.~A.,  {Scott} K.~S.,  {Wilson}
  G.~W.,  {Aretxaga} I.,  {Hughes} D.~H.,  {Almaini} O.,  {Chapin} E.~L.,
  {Chapman} S.~C.,  {Cirasuolo} M.,  {Clements} D.~L.,  {Coppin} K.~E.~K.,
  {Dunne} L.,  {Dye} S.,  {Eales} S.~A.,  {Egami} E.,    {Farrah} D. e.~a.,
  2010, MNRAS, 401, 160

\bibitem[\protect\citeauthoryear{{B{\'e}thermin}, {Dole}, {Lagache}, {Le
  Borgne} \& {Penin}}{{B{\'e}thermin} et~al.}{2011}]{bethermin_2011}
{B{\'e}thermin} M.,  {Dole} H.,  {Lagache} G.,  {Le Borgne} D.,    {Penin} A.,
  2011, A\&A, 529, A4

\bibitem[\protect\citeauthoryear{{Blain}, {Smail}, {Ivison}, {Kneib} \&
  {Frayer}}{{Blain} et~al.}{2002}]{blain_2002}
{Blain} A.~W.,  {Smail} I.,  {Ivison} R.~J.,  {Kneib} J.,    {Frayer} D.~T.,
  2002, Physics Reports, 369, 111

\bibitem[\protect\citeauthoryear{{Bonavera}, {Massardi}, {Bonaldi},
  {Gonz{\'a}lez-Nuevo}, {de Zotti} \& {Ekers}}{{Bonavera}
  et~al.}{2011}]{bonavera_2011}
{Bonavera} L.,  {Massardi} M.,  {Bonaldi} A.,  {Gonz{\'a}lez-Nuevo} J.,  {de
  Zotti} G.,    {Ekers} R.~D.,  2011, MNRAS, 416, 559

\bibitem[\protect\citeauthoryear{{Cai}, {Lapi}, {Xia}, {De Zotti}, {Negrello},
  {Gruppioni}, {Rigby}, {Castex}, {Delabrouille} \& {Danese}}{{Cai}
  et~al.}{2013}]{cai_2013}
{Cai} Z.-Y.,  {Lapi} A.,  {Xia} J.-Q.,  {De Zotti} G.,  {Negrello} M.,
  {Gruppioni} C.,  {Rigby} E.,  {Castex} G.,  {Delabrouille} J.,    {Danese}
  L.,  2013, ApJ, 768, 21

\bibitem[\protect\citeauthoryear{{Carlstrom}, {Ade}, {Aird}, {Benson}, {Bleem},
  {Busetti}, {Chang}, {Chauvin}, {Cho}, {Crawford}, {Crites}, {Dobbs},
  {Halverson}, {Heimsath}, {Holzapfel}, {Hrubes} \& {Joy}}{{Carlstrom}
  et~al.}{2011}]{carlstrom_2011}
{Carlstrom} J.~E.,  {Ade} P.~A.~R.,  {Aird} K.~A.,  {Benson} B.~A.,  {Bleem}
  L.~E.,  {Busetti} S.,  {Chang} C.~L.,  {Chauvin} E.,  {Cho} H.-M.,
  {Crawford} T.~M.,  {Crites} A.~T.,  {Dobbs} M.~A.,  {Halverson} N.~W.,
  {Heimsath} S.,  {Holzapfel} W.~L.,  {Hrubes} J.~D.,    {Joy} M. e.~a.,  2011,
  PASP, 123, 568

\bibitem[\protect\citeauthoryear{{Crawford}, {Switzer}, {Holzapfel},
  {Reichardt}, {Marrone} \& {Vieira}}{{Crawford} et~al.}{2010}]{crawford_2010}
{Crawford} T.~M.,  {Switzer} E.~R.,  {Holzapfel} W.~L.,  {Reichardt} C.~L.,
  {Marrone} D.~P.,    {Vieira} J.~D.,  2010, ApJ, 718, 513

\bibitem[\protect\citeauthoryear{{de Zotti}, {Massardi}, {Negrello} \&
  {Wall}}{{de Zotti} et~al.}{2010}]{dezotti_2010}
{de Zotti} G.,  {Massardi} M.,  {Negrello} M.,    {Wall} J.,  2010, A\&AR, 18,
  1

\bibitem[\protect\citeauthoryear{{de Zotti}, {Ricci}, {Mesa}, {Silva},
  {Mazzotta}, {Toffolatti} \& {Gonz{\'a}lez-Nuevo}}{{de Zotti}
  et~al.}{2005}]{dezotti_2005}
{de Zotti} G.,  {Ricci} R.,  {Mesa} D.,  {Silva} L.,  {Mazzotta} P.,
  {Toffolatti} L.,    {Gonz{\'a}lez-Nuevo} J.,  2005, A\&A, 431, 893

\bibitem[\protect\citeauthoryear{{Devereux} \& {Young}}{{Devereux} \&
  {Young}}{1990}]{devereux_1990}
{Devereux} N.~A.,  {Young} J.~S.,  1990, ApJ, 359, 42

\bibitem[\protect\citeauthoryear{{Devlin}, {Ade}, {Aretxaga}, {Bock}, {Chapin},
  {Griffin}, {Gundersen}, {Halpern}, {Hargrave}, {Hughes}, {Klein}, {Marsden},
  {Martin}, {Mauskopf}, {Moncelsi}, {Netterfield}, {Ngo} \& {Olmi}}{{Devlin}
  et~al.}{2009}]{devlin_2009}
{Devlin} M.~J.,  {Ade} P.~A.~R.,  {Aretxaga} I.,  {Bock} J.~J.,  {Chapin}
  E.~L.,  {Griffin} M.,  {Gundersen} J.~O.,  {Halpern} M.,  {Hargrave} P.~C.,
  {Hughes} D.~H.,  {Klein} J.,  {Marsden} G.,  {Martin} P.~G.,  {Mauskopf} P.,
  {Moncelsi} L.,  {Netterfield} C.~B.,  {Ngo} H.,    {Olmi} L. e.~a.,  2009,
  Nature, 458, 737

\bibitem[\protect\citeauthoryear{{Dole}, {Lagache}, {Puget}, {Caputi},
  {Fern{\'a}ndez-Conde}, {Le Floc'h}, {Papovich}, {P{\'e}rez-Gonz{\'a}lez},
  {Rieke} \& {Blaylock}}{{Dole} et~al.}{2006}]{dole_2006}
{Dole} H.,  {Lagache} G.,  {Puget} J.,  {Caputi} K.~I.,  {Fern{\'a}ndez-Conde}
  N.,  {Le Floc'h} E.,  {Papovich} C.,  {P{\'e}rez-Gonz{\'a}lez} P.~G.,
  {Rieke} G.~H.,    {Blaylock} M.,  2006, A\&A, 451, 417

\bibitem[\protect\citeauthoryear{{Draine}}{{Draine}}{2003}]{draine_2003}
{Draine} B.~T.,  2003, ARA\&A, 41, 241

\bibitem[\protect\citeauthoryear{{Draine} \& {Lee}}{{Draine} \&
  {Lee}}{1984}]{draine_1984}
{Draine} B.~T.,  {Lee} H.~M.,  1984, ApJ, 285, 89

\bibitem[\protect\citeauthoryear{{Dunkley}, {Hlozek}, {Sievers}, {Acquaviva},
  {Ade}, {Aguirre}, {Amiri}, {Appel}, {Barrientos}, {Battistelli}, {Bond},
  {Brown} \& {Burger}}{{Dunkley} et~al.}{2011}]{dunkley_2011}
{Dunkley} J.,  {Hlozek} R.,  {Sievers} J.,  {Acquaviva} V.,  {Ade} P.~A.~R.,
  {Aguirre} P.,  {Amiri} M.,  {Appel} J.~W.,  {Barrientos} L.~F.,
  {Battistelli} E.~S.,  {Bond} J.~R.,  {Brown} B.,    {Burger} B. e.~a.,  2011,
  ApJ, 739, 52

\bibitem[\protect\citeauthoryear{{D{\"u}nner}, {Hasselfield}, {Marriage},
  {Sievers}, {Acquaviva}, {Addison}, {Ade}, {Aguirre}, {Amiri}, {Appel},
  {Barrientos}, {Battistelli}, {Bond}, {Brown}, {Burger}, {Calabrese} \&
  {Chervenak}}{{D{\"u}nner} et~al.}{2013}]{dunner_2013}
{D{\"u}nner} R.,  {Hasselfield} M.,  {Marriage} T.~A.,  {Sievers} J.,
  {Acquaviva} V.,  {Addison} G.~E.,  {Ade} P.~A.~R.,  {Aguirre} P.,  {Amiri}
  M.,  {Appel} J.~W.,  {Barrientos} L.~F.,  {Battistelli} E.~S.,  {Bond} J.~R.,
   {Brown} B.,  {Burger} B.,  {Calabrese} E.,    {Chervenak} J. e.~a.,  2013,
  ApJ, 762, 10

\bibitem[\protect\citeauthoryear{{Ghisellini} \& {Tavecchio}}{{Ghisellini} \&
  {Tavecchio}}{2008}]{ghisellini_2008}
{Ghisellini} G.,  {Tavecchio} F.,  2008, MNRAS, 387, 1669

\bibitem[\protect\citeauthoryear{{Goldin}, {Kowitt}, {Cheng}, {Cottingham},
  {Fixsen}, {Inman}, {Meyer}, {Puchalla}, {Ruhl} \& {Silverberg}}{{Goldin}
  et~al.}{1997}]{goldin_1997}
{Goldin} A.~B.,  {Kowitt} M.~S.,  {Cheng} E.~S.,  {Cottingham} D.~A.,  {Fixsen}
  D.~J.,  {Inman} C.~A.,  {Meyer} S.~S.,  {Puchalla} J.~L.,  {Ruhl} J.~E.,
  {Silverberg} R.~F.,  1997, ApJL, 488, L161

\bibitem[\protect\citeauthoryear{{Griffin} \& {Orton}}{{Griffin} \&
  {Orton}}{1993}]{griffin_orton_1993}
{Griffin} M.~J.,  {Orton} G.~S.,  1993, Icarus, 105, 537

\bibitem[\protect\citeauthoryear{{Hajian}, {Acquaviva}, {Ade}, {Aguirre},
  {Amiri}, {Appel}, {Barrientos}, {Battistelli}, {Bond}, {Brown}, {Burger},
  {Chervenak}, {Das}, {Devlin}, {Dicker}, {Bertrand Doriese}, {Dunkley} \&
  {D{\"u}nner}}{{Hajian} et~al.}{2011}]{hajian_2010}
{Hajian} A.,  {Acquaviva} V.,  {Ade} P.~A.~R.,  {Aguirre} P.,  {Amiri} M.,
  {Appel} J.~W.,  {Barrientos} L.~F.,  {Battistelli} E.~S.,  {Bond} J.~R.,
  {Brown} B.,  {Burger} B.,  {Chervenak} J.,  {Das} S.,  {Devlin} M.~J.,
  {Dicker} S.~R.,  {Bertrand Doriese} W.,  {Dunkley} J.,    {D{\"u}nner} R.
  e.~a.,  2011, ApJ, 740, 86

\bibitem[\protect\citeauthoryear{{Hall}, {Keisler}, {Knox}, {Reichardt}, {Ade},
  {Aird}, {Benson}, {Bleem}, {Carlstrom}, {Chang}, {Cho}, {Crawford} \&
  {Crites}}{{Hall} et~al.}{2010}]{hall_2010}
{Hall} N.~R.,  {Keisler} R.,  {Knox} L.,  {Reichardt} C.~L.,  {Ade} P.~A.~R.,
  {Aird} K.~A.,  {Benson} B.~A.,  {Bleem} L.~E.,  {Carlstrom} J.~E.,  {Chang}
  C.~L.,  {Cho} H.-M.,  {Crawford} T.~M.,    {Crites} A.~T. e.~a.,  2010, ApJ,
  718, 632

\bibitem[\protect\citeauthoryear{{Hasselfield}, {Moodley}, {Bond}, {Das},
  {Devlin}, {Dunkley}, {Dunner}, {Fowler}, {Gallardo}, {Gralla}, {Hajian},
  {Halpern}, {Hincks}, {Marriage}, {Marsden}, {Niemack}, {Nolta} \&
  {Page}}{{Hasselfield} et~al.}{2013}]{hasselfield_2013}
{Hasselfield} M.,  {Moodley} K.,  {Bond} J.~R.,  {Das} S.,  {Devlin} M.~J.,
  {Dunkley} J.,  {Dunner} R.,  {Fowler} J.~W.,  {Gallardo} P.,  {Gralla} M.~B.,
   {Hajian} A.,  {Halpern} M.,  {Hincks} A.~D.,  {Marriage} T.~A.,  {Marsden}
  D.,  {Niemack} M.~D.,  {Nolta} M.~R.,    {Page} L.~A. e.~a.,  2013, ArXiv
  e-prints

\bibitem[\protect\citeauthoryear{{Helou}, {Khan}, {Malek} \& {Boehmer}}{{Helou}
  et~al.}{1988}]{helou_1988}
{Helou} G.,  {Khan} I.~R.,  {Malek} L.,    {Boehmer} L.,  1988, ApJS, 68, 151

\bibitem[\protect\citeauthoryear{{Hezaveh}, {Marrone}, {Fassnacht}, {Spilker},
  {Vieira}, {Aguirre}, {Aird}, {Aravena}, {Ashby}, {Bayliss}, {Benson},
  {Bleem}, {Bothwell}, {Brodwin}, {Carlstrom}, {Chang} \& {Chapman}}{{Hezaveh}
  et~al.}{2013}]{hezaveh_2013}
{Hezaveh} Y.~D.,  {Marrone} D.~P.,  {Fassnacht} C.~D.,  {Spilker} J.~S.,
  {Vieira} J.~D.,  {Aguirre} J.~E.,  {Aird} K.~A.,  {Aravena} M.,  {Ashby}
  M.~L.~N.,  {Bayliss} M.,  {Benson} B.~A.,  {Bleem} L.~E.,  {Bothwell} M.,
  {Brodwin} M.,  {Carlstrom} J.~E.,  {Chang} C.~L.,    {Chapman} S.~C. e.~a.,
  2013, ApJ, 767, 132

\bibitem[\protect\citeauthoryear{{Hincks}, {Acquaviva}, {Ade}, {Aguirre},
  {Amiri}, {Appel}, {Barrientos}, {Battistelli}, {Bond}, {Brown}, {Burger},
  {Chervenak}, {Das}, {Devlin}, {Dicker}, {Doriese}, {Dunkley} \&
  {D{\"u}nner}}{{Hincks} et~al.}{2010}]{hincks_2010}
{Hincks} A.~D.,  {Acquaviva} V.,  {Ade} P.~A.~R.,  {Aguirre} P.,  {Amiri} M.,
  {Appel} J.~W.,  {Barrientos} L.~F.,  {Battistelli} E.~S.,  {Bond} J.~R.,
  {Brown} B.,  {Burger} B.,  {Chervenak} J.,  {Das} S.,  {Devlin} M.~J.,
  {Dicker} S.~R.,  {Doriese} W.~B.,  {Dunkley} J.,    {D{\"u}nner} R. e.~a.,
  2010, ApJS, 191, 423

\bibitem[\protect\citeauthoryear{{Holland}, {Bintley}, {Chapin},
  {Chrysostomou}, {Davis}, {Dempsey}, {Duncan}, {Fich}, {Friberg}, {Halpern},
  {Irwin}, {Jenness} \& {Kelly}}{{Holland} et~al.}{2013}]{holland_2013}
{Holland} W.~S.,  {Bintley} D.,  {Chapin} E.~L.,  {Chrysostomou} A.,  {Davis}
  G.~R.,  {Dempsey} J.~T.,  {Duncan} W.~D.,  {Fich} M.,  {Friberg} P.,
  {Halpern} M.,  {Irwin} K.~D.,  {Jenness} T.,    {Kelly} B.~D. e.~a.,  2013,
  MNRAS, 430, 2513

\bibitem[\protect\citeauthoryear{{Holland}, {Robson}, {Gear}, {Cunningham},
  {Lightfoot}, {Jenness}, {Ivison}, {Stevens}, {Ade}, {Griffin}, {Duncan},
  {Murphy} \& {Naylor}}{{Holland} et~al.}{1999}]{holland_1999}
{Holland} W.~S.,  {Robson} E.~I.,  {Gear} W.~K.,  {Cunningham} C.~R.,
  {Lightfoot} J.~F.,  {Jenness} T.,  {Ivison} R.~J.,  {Stevens} J.~A.,  {Ade}
  P.~A.~R.,  {Griffin} M.~J.,  {Duncan} W.~D.,  {Murphy} J.~A.,    {Naylor}
  D.~A.,  1999, MNRAS, 303, 659

\bibitem[\protect\citeauthoryear{{Ikarashi}, {Kohno}, {Aguirre}, {Aretxaga},
  {Arumugam}, {Austermann}, {Bock}, {Bradford}, {Cirasuolo}, {Earle}, {Ezawa},
  {Furusawa}, {Furusawa}, {Glenn}, {Hatsukade}, {Hughes}, {Iono} \&
  {Ivison}}{{Ikarashi} et~al.}{2011}]{ikarashi_2010}
{Ikarashi} S.,  {Kohno} K.,  {Aguirre} J.~E.,  {Aretxaga} I.,  {Arumugam} V.,
  {Austermann} J.~E.,  {Bock} J.~J.,  {Bradford} C.~M.,  {Cirasuolo} M.,
  {Earle} L.,  {Ezawa} H.,  {Furusawa} H.,  {Furusawa} J.,  {Glenn} J.,
  {Hatsukade} B.,  {Hughes} D.~H.,  {Iono} D.,    {Ivison} R.~J. e.~a.,  2011,
  MNRAS, 415, 3081

\bibitem[\protect\citeauthoryear{{Lamarre}, {Puget}, {Ade}, {Bouchet}, {Guyot},
  {Lange}, {Pajot}, {Arondel}, {Benabed}, {Beney}, {Beno{\^i}t}, {Bernard},
  {Bhatia}, {Blanc}, {Bock}, {Br{\'e}elle}, {Bradshaw} \& {Camus}}{{Lamarre}
  et~al.}{2010}]{lamarre_2010}
{Lamarre} J.-M.,  {Puget} J.-L.,  {Ade} P.~A.~R.,  {Bouchet} F.,  {Guyot} G.,
  {Lange} A.~E.,  {Pajot} F.,  {Arondel} A.,  {Benabed} K.,  {Beney} J.-L.,
  {Beno{\^i}t} A.,  {Bernard} J.-P.,  {Bhatia} R.,  {Blanc} Y.,  {Bock} J.~J.,
  {Br{\'e}elle} E.,  {Bradshaw} T.~W.,    {Camus} P. e.~a.,  2010, A\&A, 520,
  A9

\bibitem[\protect\citeauthoryear{{Lima}, {Jain}, {Devlin} \& {Aguirre}}{{Lima}
  et~al.}{2010}]{lima_2010}
{Lima} M.,  {Jain} B.,  {Devlin} M.,    {Aguirre} J.,  2010, ApJL, 717, L31

\bibitem[\protect\citeauthoryear{{Lupu}, {Scott}, {Aguirre}, {Aretxaga},
  {Auld}, {Barton}, {Beelen}, {Bertoldi}, {Bock}, {Bonfield}, {Bradford},
  {Buttiglione}, {Cava}, {Clements}, {Cooke}, {Cooray}, {Dannerbauer},
  {Dariush} \& {De Zotti}}{{Lupu} et~al.}{2012}]{lupu_2012}
{Lupu} R.~E.,  {Scott} K.~S.,  {Aguirre} J.~E.,  {Aretxaga} I.,  {Auld} R.,
  {Barton} E.,  {Beelen} A.,  {Bertoldi} F.,  {Bock} J.~J.,  {Bonfield} D.,
  {Bradford} C.~M.,  {Buttiglione} S.,  {Cava} A.,  {Clements} D.~L.,  {Cooke}
  J.,  {Cooray} A.,  {Dannerbauer} H.,  {Dariush} A.,    {De Zotti} G. e.~a.,
  2012, ApJ, 757, 135

\bibitem[\protect\citeauthoryear{{Marriage}, {Baptiste Juin}, {Lin}, {Marsden},
  {Nolta}, {Partridge}, {Ade}, {Aguirre}, {Amiri}, {Appel}, {Barrientos},
  {Battistelli}, {Bond}, {Brown}, {Burger}, {Chervenak}, {Das} \&
  {Devlin}}{{Marriage} et~al.}{2011}]{marriage_2011a}
{Marriage} T.~A.,  {Baptiste Juin} J.,  {Lin} Y.-T.,  {Marsden} D.,  {Nolta}
  M.~R.,  {Partridge} B.,  {Ade} P.~A.~R.,  {Aguirre} P.,  {Amiri} M.,  {Appel}
  J.~W.,  {Barrientos} L.~F.,  {Battistelli} E.~S.,  {Bond} J.~R.,  {Brown} B.,
   {Burger} B.,  {Chervenak} J.,  {Das} S.,    {Devlin} M.~J. e.~a.,  2011,
  ApJ, 731, 100

\bibitem[\protect\citeauthoryear{{Marscher} \& {Gear}}{{Marscher} \&
  {Gear}}{1985}]{marscher_1985}
{Marscher} A.~P.,  {Gear} W.~K.,  1985, ApJ, 298, 114

\bibitem[\protect\citeauthoryear{{Massardi}, {Bonaldi}, {Bonavera},
  {L{\'o}pez-Caniego}, {de Zotti} \& {Ekers}}{{Massardi}
  et~al.}{2011}]{massardi_2011b}
{Massardi} M.,  {Bonaldi} A.,  {Bonavera} L.,  {L{\'o}pez-Caniego} M.,  {de
  Zotti} G.,    {Ekers} R.~D.,  2011, MNRAS, pp 878--+

\bibitem[\protect\citeauthoryear{{Massardi, M. et al.}}{{Massardi, M. et
  al.}}{2011}]{massardi_2011a}
{Massardi, M. et al.} 2011, MNRAS, 412, 318

\bibitem[\protect\citeauthoryear{{Mauch}, {Murphy}, {Buttery}, {Curran},
  {Hunstead}, {Piestrzynski}, {Robertson} \& {Sadler}}{{Mauch}
  et~al.}{2003}]{mauch_2003}
{Mauch} T.,  {Murphy} T.,  {Buttery} H.~J.,  {Curran} J.,  {Hunstead} R.~W.,
  {Piestrzynski} B.,  {Robertson} J.~G.,    {Sadler} E.~M.,  2003, MNRAS, 342,
  1117

\bibitem[\protect\citeauthoryear{{Mocanu}, {Crawford}, {Vieira}, {Aird},
  {Aravena}, {Austermann}, {Benson}, {B{\'e}thermin}, {Bleem}, {Bothwell},
  {Carlstrom}, {Chang} \& {Chapman}}{{Mocanu} et~al.}{2013}]{mocanu_2013}
{Mocanu} L.~M.,  {Crawford} T.~M.,  {Vieira} J.~D.,  {Aird} K.~A.,  {Aravena}
  M.,  {Austermann} J.~E.,  {Benson} B.~A.,  {B{\'e}thermin} M.,  {Bleem}
  L.~E.,  {Bothwell} M.,  {Carlstrom} J.~E.,  {Chang} C.~L.,    {Chapman} S.
  e.~a.,  2013, ArXiv e-prints

\bibitem[\protect\citeauthoryear{{Murphy}, {Sadler}, {Ekers}, {Massardi},
  {Hancock}, {Mahony}, {Ricci}, {Burke-Spolaor}, {Calabretta}, {Chhetri}, {de
  Zotti}, {Edwards}, {Ekers}, {Jackson}, {Kesteven}, {Lindley} \&
  {Newton-McGee}}{{Murphy} et~al.}{2010}]{murphy_2010}
{Murphy} T.,  {Sadler} E.~M.,  {Ekers} R.~D.,  {Massardi} M.,  {Hancock} P.~J.,
   {Mahony} E.,  {Ricci} R.,  {Burke-Spolaor} S.,  {Calabretta} M.,  {Chhetri}
  R.,  {de Zotti} G.,  {Edwards} P.~G.,  {Ekers} J.~A.,  {Jackson} C.~A.,
  {Kesteven} M.~J.,  {Lindley} E.,    {Newton-McGee} K. e.~a.,  2010, MNRAS,
  402, 2403

\bibitem[\protect\citeauthoryear{{Negrello}, {Hopwood}, {De Zotti}, {Cooray},
  {Verma}, {Bock}, {Frayer}, {Gurwell}, {Omont}, {Neri}, {Dannerbauer},
  {Leeuw}, {Barton}, {Cooke}, {Kim}, {da Cunha}, {Rodighiero}, {Cox} \&
  {Bonfield}}{{Negrello} et~al.}{2010}]{negrello_2010}
{Negrello} M.,  {Hopwood} R.,  {De Zotti} G.,  {Cooray} A.,  {Verma} A.,
  {Bock} J.,  {Frayer} D.~T.,  {Gurwell} M.~A.,  {Omont} A.,  {Neri} R.,
  {Dannerbauer} H.,  {Leeuw} L.~L.,  {Barton} E.,  {Cooke} J.,  {Kim} S.,  {da
  Cunha} E.,  {Rodighiero} G.,  {Cox} P.,    {Bonfield} D.~G. e.~a.,  2010,
  Science, 330, 800

\bibitem[\protect\citeauthoryear{{Negrello}, {Perrotta}, {Gonz{\'a}lez-Nuevo},
  {Silva}, {de Zotti}, {Granato}, {Baccigalupi} \& {Danese}}{{Negrello}
  et~al.}{2007}]{negrello_2007}
{Negrello} M.,  {Perrotta} F.,  {Gonz{\'a}lez-Nuevo} J.,  {Silva} L.,  {de
  Zotti} G.,  {Granato} G.~L.,  {Baccigalupi} C.,    {Danese} L.,  2007, MNRAS,
  377, 1557

\bibitem[\protect\citeauthoryear{{Nieppola}, {Valtaoja}, {Tornikoski},
  {Hovatta} \& {Kotiranta}}{{Nieppola} et~al.}{2008}]{nieppola_2008}
{Nieppola} E.,  {Valtaoja} E.,  {Tornikoski} M.,  {Hovatta} T.,    {Kotiranta}
  M.,  2008, A\&A, 488, 867

\bibitem[\protect\citeauthoryear{{Pascale}, {Ade}, {Bock}, {Chapin}, {Chung},
  {Devlin}, {Dicker}, {Griffin}, {Gundersen}, {Halpern}, {Hargrave}, {Hughes}
  \& {Klein}}{{Pascale} et~al.}{2008}]{pascale_2008}
{Pascale} E.,  {Ade} P.~A.~R.,  {Bock} J.~J.,  {Chapin} E.~L.,  {Chung} J.,
  {Devlin} M.~J.,  {Dicker} S.,  {Griffin} M.,  {Gundersen} J.~O.,  {Halpern}
  M.,  {Hargrave} P.~C.,  {Hughes} D.~H.,    {Klein} J. e.~a.,  2008, ApJ, 681,
  400

\bibitem[\protect\citeauthoryear{{Pilbratt}, {Riedinger}, {Passvogel}, {Crone},
  {Doyle}, {Gageur}, {Heras}, {Jewell}, {Metcalfe}, {Ott} \&
  {Schmidt}}{{Pilbratt} et~al.}{2010}]{pilbratt_2010}
{Pilbratt} G.~L.,  {Riedinger} J.~R.,  {Passvogel} T.,  {Crone} G.,  {Doyle}
  D.,  {Gageur} U.,  {Heras} A.~M.,  {Jewell} C.,  {Metcalfe} L.,  {Ott} S.,
  {Schmidt} M.,  2010, A\&A, 518, L1

\bibitem[\protect\citeauthoryear{{Planck Collaboration VII}}{{Planck
  Collaboration VII}}{2013}]{planck7_2012}
{Planck Collaboration VII} 2013, A\&A, 550, A133

\bibitem[\protect\citeauthoryear{{Planck Collaboration XIII}}{{Planck
  Collaboration XIII}}{2011}]{planck13_2011}
{Planck Collaboration XIII} 2011, A\&A, 536, A13

\bibitem[\protect\citeauthoryear{{Planck Collaboration XV}}{{Planck
  Collaboration XV}}{2011}]{planck15_2011}
{Planck Collaboration XV} 2011, A\&A, 536, A15

\bibitem[\protect\citeauthoryear{{Planck Collaboration XVI}}{{Planck
  Collaboration XVI}}{2011}]{planck17_2011}
{Planck Collaboration XVI} 2011, A\&A, 536, A16

\bibitem[\protect\citeauthoryear{{Planck Collaboration XXVIII}}{{Planck
  Collaboration XXVIII}}{2013}]{planck28_2013}
{Planck Collaboration XXVIII} 2013, ArXiv e-prints

\bibitem[\protect\citeauthoryear{Press, Teukolsky, Vetterling \&
  Flannery}{Press et~al.}{2007}]{press_2007}
Press W.~H.,  Teukolsky S.~A.,  Vetterling W.~T.,    Flannery B.~P.,  2007,
  Numerical recipes in C (3rd ed.): the art of scientific computing.
Cambridge University Press, New York, NY, USA

\bibitem[\protect\citeauthoryear{{Rex}, {Rawle}, {Egami},
  {P{\'e}rez-Gonz{\'a}lez}, {Zemcov}, {Aretxaga}, {Chung}, {Fadda}, {Gonzalez},
  {Hughes}, {Horellou}, {Johansson}, {Kneib}, {Richard}, {Altieri}, {Fiedler},
  {Pereira} \& {Rieke}}{{Rex} et~al.}{2010}]{rex_2010}
{Rex} M.,  {Rawle} T.~D.,  {Egami} E.,  {P{\'e}rez-Gonz{\'a}lez} P.~G.,
  {Zemcov} M.,  {Aretxaga} I.,  {Chung} S.~M.,  {Fadda} D.,  {Gonzalez} A.~H.,
  {Hughes} D.~H.,  {Horellou} C.,  {Johansson} D.,  {Kneib} J.-P.,  {Richard}
  J.,  {Altieri} B.,  {Fiedler} A.~K.,  {Pereira} M.~J.,    {Rieke} G.~H.
  e.~a.,  2010, A\&A, 518, L13

\bibitem[\protect\citeauthoryear{{Sambruna}, {Donato}, {Ajello}, {Maraschi},
  {Tueller}, {Baumgartner}, {Skinner}, {Markwardt}, {Barthelmy}, {Gehrels} \&
  {Mushotzky}}{{Sambruna} et~al.}{2010}]{sambruna_2010}
{Sambruna} R.~M.,  {Donato} D.,  {Ajello} M.,  {Maraschi} L.,  {Tueller} J.,
  {Baumgartner} W.,  {Skinner} G.,  {Markwardt} C.,  {Barthelmy} S.,  {Gehrels}
  N.,    {Mushotzky} R.~F.,  2010, ApJ, 710, 24

\bibitem[\protect\citeauthoryear{{Sault}, {Teuben} \& {Wright}}{{Sault}
  et~al.}{1995}]{sault_1995}
{Sault} R.~J.,  {Teuben} P.~J.,    {Wright} M.~C.~H.,  1995, in {R.~A.~Shaw,
  H.~E.~Payne, \& J.~J.~E.~Hayes} ed., Astronomical Data Analysis Software and
  Systems IV Vol.~77 of Astronomical Society of the Pacific Conference Series,
  {A Retrospective View of MIRIAD}.
pp 433--+

\bibitem[\protect\citeauthoryear{{Scott}, {Lupu}, {Aguirre}, {Auld}, {Aussel},
  {Baker}, {Beelen}, {Bock}, {Bradford}, {Brisbin}, {Burgarella}, {Carpenter},
  {Chanial}, {Chapman}, {Clements}, {Conley}, {Cooray} \& {Cox}}{{Scott}
  et~al.}{2011}]{scott_2011}
{Scott} K.~S.,  {Lupu} R.~E.,  {Aguirre} J.~E.,  {Auld} R.,  {Aussel} H.,
  {Baker} A.~J.,  {Beelen} A.,  {Bock} J.,  {Bradford} C.~M.,  {Brisbin} D.,
  {Burgarella} D.,  {Carpenter} J.~M.,  {Chanial} P.,  {Chapman} S.~C.,
  {Clements} D.~L.,  {Conley} A.,  {Cooray} A.,    {Cox} P. e.~a.,  2011, ApJ,
  733, 29

\bibitem[\protect\citeauthoryear{{Sehgal}, {Bode}, {Das},
  {Hernandez-Monteagudo}, {Huffenberger}, {Lin}, {Ostriker} \& {Trac}}{{Sehgal}
  et~al.}{2010}]{sehgal_2010}
{Sehgal} N.,  {Bode} P.,  {Das} S.,  {Hernandez-Monteagudo} C.,  {Huffenberger}
  K.,  {Lin} Y.,  {Ostriker} J.~P.,    {Trac} H.,  2010, ApJ, 709, 920

\bibitem[\protect\citeauthoryear{{Siringo}, {Kreysa}, {Kov{\'a}cs}, {Schuller},
  {Wei{\ss}}, {Esch}, {Gem{\"u}nd}, {Jethava}, {Lundershausen}, {Colin},
  {G{\"u}sten}, {Menten}, {Beelen}, {Bertoldi}, {Beeman} \& {Haller}}{{Siringo}
  et~al.}{2009}]{siringo_2009}
{Siringo} G.,  {Kreysa} E.,  {Kov{\'a}cs} A.,  {Schuller} F.,  {Wei{\ss}} A.,
  {Esch} W.,  {Gem{\"u}nd} H.-P.,  {Jethava} N.,  {Lundershausen} G.,  {Colin}
  A.,  {G{\"u}sten} R.,  {Menten} K.~M.,  {Beelen} A.,  {Bertoldi} F.,
  {Beeman} J.~W.,    {Haller} E.~E.,  2009, A\&A, 497, 945

\bibitem[\protect\citeauthoryear{{Swetz}, {Ade}, {Amiri}, {Appel},
  {Battistelli}, {Burger}, {Chervenak}, {Devlin}, {Dicker}, {Doriese},
  {D{\"u}nner}, {Essinger-Hileman}, {Fisher}, {Fowler}, {Halpern} \&
  {Hasselfield}}{{Swetz} et~al.}{2011}]{swetz_2011}
{Swetz} D.~S.,  {Ade} P.~A.~R.,  {Amiri} M.,  {Appel} J.~W.,  {Battistelli}
  E.~S.,  {Burger} B.,  {Chervenak} J.,  {Devlin} M.~J.,  {Dicker} S.~R.,
  {Doriese} W.~B.,  {D{\"u}nner} R.,  {Essinger-Hileman} T.,  {Fisher} R.~P.,
  {Fowler} J.~W.,  {Halpern} M.,    {Hasselfield} M. e.~a.,  2011, ApJS, 194,
  41

\bibitem[\protect\citeauthoryear{{Tegmark}}{{Tegmark}}{1997}]{tegmark_1997}
{Tegmark} M.,  1997, ApJL, 480, L87+

\bibitem[\protect\citeauthoryear{{Thomas}, {Maraston}, {Bender} \& {Mendes de
  Oliveira}}{{Thomas} et~al.}{2005}]{thomas_2005}
{Thomas} D.,  {Maraston} C.,  {Bender} R.,    {Mendes de Oliveira} C.,  2005,
  ApJ, 621, 673

\bibitem[\protect\citeauthoryear{{Thompson}, {Moran} \& {Swenson}}{{Thompson}
  et~al.}{1986}]{thompson_1986}
{Thompson} A.~R.,  {Moran} J.~M.,    {Swenson} G.~W.,  1986, {Interferometry
  and synthesis in radio astronomy}

\bibitem[\protect\citeauthoryear{{Toffolatti}, {Argueso Gomez}, {de Zotti},
  {Mazzei}, {Franceschini}, {Danese} \& {Burigana}}{{Toffolatti}
  et~al.}{1998}]{toffolatti_1998}
{Toffolatti} L.,  {Argueso Gomez} F.,  {de Zotti} G.,  {Mazzei} P.,
  {Franceschini} A.,  {Danese} L.,    {Burigana} C.,  1998, MNRAS, 297, 117

\bibitem[\protect\citeauthoryear{{Tucci}, {Toffolatti}, {de Zotti} \&
  {Mart{\'{\i}}nez-Gonz{\'a}lez}}{{Tucci} et~al.}{2011}]{tucci_2011}
{Tucci} M.,  {Toffolatti} L.,  {de Zotti} G.,    {Mart{\'{\i}}nez-Gonz{\'a}lez}
  E.,  2011, A\&A, 533, A57

\bibitem[\protect\citeauthoryear{{Ulich}}{{Ulich}}{1981}]{ulich_1981}
{Ulich} B.~L.,  1981, AJ, 86, 1619

\bibitem[\protect\citeauthoryear{{Valtaoja}, {Terasranta}, {Urpo}, {Nesterov},
  {Lainela} \& {Valtonen}}{{Valtaoja} et~al.}{1992}]{valtaoja_1992}
{Valtaoja} E.,  {Terasranta} H.,  {Urpo} S.,  {Nesterov} N.~S.,  {Lainela} M.,
    {Valtonen} M.,  1992, A\&A, 254, 71

\bibitem[\protect\citeauthoryear{{Vanderlinde}, {Crawford}, {de Haan},
  {Dudley}, {Shaw}, {Ade}, {Aird}, {Benson}, {Bleem}, {Brodwin}, {Carlstrom},
  {Chang}, {Crites}, {Desai}, {Dobbs}, {Foley}, {George} \&
  {Gladders}}{{Vanderlinde} et~al.}{2010}]{vanderlinde_2010}
{Vanderlinde} K.,  {Crawford} T.~M.,  {de Haan} T.,  {Dudley} J.~P.,  {Shaw}
  L.,  {Ade} P.~A.~R.,  {Aird} K.~A.,  {Benson} B.~A.,  {Bleem} L.~E.,
  {Brodwin} M.,  {Carlstrom} J.~E.,  {Chang} C.~L.,  {Crites} A.~T.,  {Desai}
  S.,  {Dobbs} M.~A.,  {Foley} R.~J.,  {George} E.~M.,    {Gladders} M.~D.
  e.~a.,  2010, ApJ, 722, 1180

\bibitem[\protect\citeauthoryear{{Vieira}, {Crawford}, {Switzer}, {Ade},
  {Aird}, {Ashby}, {Benson}, {Bleem}, {Brodwin}, {Carlstrom}, {Chang}, {Cho},
  {Crites}, {de Haan}, {Dobbs}, {Everett}, {George} \& {Gladders}}{{Vieira}
  et~al.}{2010}]{vieira_2010}
{Vieira} J.~D.,  {Crawford} T.~M.,  {Switzer} E.~R.,  {Ade} P.~A.~R.,  {Aird}
  K.~A.,  {Ashby} M.~L.~N.,  {Benson} B.~A.,  {Bleem} L.~E.,  {Brodwin} M.,
  {Carlstrom} J.~E.,  {Chang} C.~L.,  {Cho} H.-M.,  {Crites} A.~T.,  {de Haan}
  T.,  {Dobbs} M.~A.,  {Everett} W.,  {George} E.~M.,    {Gladders} M. e.~a.,
  2010, ApJ, 719, 763

\bibitem[\protect\citeauthoryear{{Vieira}, {Marrone}, {Chapman}, {De Breuck},
  {Hezaveh}, {Wei{$\beta$}}, {Aguirre}, {Aird}, {Aravena}, {Ashby}, {Bayliss},
  {Benson}, {Biggs}, {Bleem}, {Bock}, {Bothwell}, {Bradford} \&
  {Brodwin}}{{Vieira} et~al.}{2013}]{vieira_2013}
{Vieira} J.~D.,  {Marrone} D.~P.,  {Chapman} S.~C.,  {De Breuck} C.,  {Hezaveh}
  Y.~D.,  {Wei{$\beta$}} A.,  {Aguirre} J.~E.,  {Aird} K.~A.,  {Aravena} M.,
  {Ashby} M.~L.~N.,  {Bayliss} M.,  {Benson} B.~A.,  {Biggs} A.~D.,  {Bleem}
  L.~E.,  {Bock} J.~J.,  {Bothwell} M.,  {Bradford} C.~M.,    {Brodwin} M.
  e.~a.,  2013, Nature, 495, 344

\bibitem[\protect\citeauthoryear{{Viero}, {Ade}, {Bock}, {Chapin}, {Devlin},
  {Griffin}, {Gundersen}, {Halpern}, {Hargrave}, {Hughes}, {Klein},
  {MacTavish}, {Marsden}, {Martin}, {Mauskopf}, {Moncelsi}, {Negrello} \&
  {Netterfield}}{{Viero} et~al.}{2009}]{viero_2009}
{Viero} M.~P.,  {Ade} P.~A.~R.,  {Bock} J.~J.,  {Chapin} E.~L.,  {Devlin}
  M.~J.,  {Griffin} M.,  {Gundersen} J.~O.,  {Halpern} M.,  {Hargrave} P.~C.,
  {Hughes} D.~H.,  {Klein} J.,  {MacTavish} C.~J.,  {Marsden} G.,  {Martin}
  P.~G.,  {Mauskopf} P.,  {Moncelsi} L.,  {Negrello} M.,    {Netterfield} C.~B.
  e.~a.,  2009, ApJ, 707, 1766

\bibitem[\protect\citeauthoryear{{Wardlow}, {Cooray}, {De Bernardis},
  {Amblard}, {Arumugam}, {Aussel}, {Baker}, {B{\'e}thermin}, {Blundell},
  {Bock}, {Boselli} \& {Bridge}}{{Wardlow} et~al.}{2013}]{wardlow_2013}
{Wardlow} J.~L.,  {Cooray} A.,  {De Bernardis} F.,  {Amblard} A.,  {Arumugam}
  V.,  {Aussel} H.,  {Baker} A.~J.,  {B{\'e}thermin} M.,  {Blundell} R.,
  {Bock} J.,  {Boselli} A.,    {Bridge} C. e.~a.,  2013, ApJ, 762, 59

\bibitem[\protect\citeauthoryear{{Weiland}, {Odegard}, {Hill}, {Wollack},
  {Hinshaw}, {Greason}, {Jarosik}, {Page}, {Bennett}, {Dunkley}, {Gold},
  {Halpern}, {Kogut}, {Komatsu}, {Larson}, {Limon}, {Meyer}, {Nolta}, {Smith}
  \& {Spergel}}{{Weiland} et~al.}{2011}]{weiland_2011}
{Weiland} J.~L.,  {Odegard} N.,  {Hill} R.~S.,  {Wollack} E.,  {Hinshaw} G.,
  {Greason} M.~R.,  {Jarosik} N.,  {Page} L.,  {Bennett} C.~L.,  {Dunkley} J.,
  {Gold} B.,  {Halpern} M.,  {Kogut} A.,  {Komatsu} E.,  {Larson} D.,  {Limon}
  M.,  {Meyer} S.~S.,  {Nolta} M.~R.,  {Smith} K.~M.,    {Spergel} D.~N. e.~a.,
   2011, ApJS, 192, 19

\bibitem[\protect\citeauthoryear{{Wei{\ss}}, {De Breuck}, {Marrone}, {Vieira},
  {Aguirre}, {Aird}, {Aravena}, {Ashby}, {Bayliss}, {Benson}, {B{\'e}thermin},
  {Biggs}, {Bleem}, {Bock}, {Bothwell}, {Bradford} \& {Brodwin}}{{Wei{\ss}}
  et~al.}{2013}]{weil_2013}
{Wei{\ss}} A.,  {De Breuck} C.,  {Marrone} D.~P.,  {Vieira} J.~D.,  {Aguirre}
  J.~E.,  {Aird} K.~A.,  {Aravena} M.,  {Ashby} M.~L.~N.,  {Bayliss} M.,
  {Benson} B.~A.,  {B{\'e}thermin} M.,  {Biggs} A.~D.,  {Bleem} L.~E.,  {Bock}
  J.~J.,  {Bothwell} M.,  {Bradford} C.~M.,    {Brodwin} M. e.~a.,  2013, ApJ,
  767, 88

\bibitem[\protect\citeauthoryear{{Wei{\ss}}, {Kov{\'a}cs}, {Coppin}, {Greve},
  {Walter}, {Smail}, {Dunlop}, {Knudsen}, {Alexander}, {Bertoldi}, {Brandt},
  {Chapman}, {Cox}, {Dannerbauer}, {De Breuck}, {Gawiser}, {Ivison} \&
  {Lutz}}{{Wei{\ss}} et~al.}{2009}]{weil_2009}
{Wei{\ss}} A.,  {Kov{\'a}cs} A.,  {Coppin} K.,  {Greve} T.~R.,  {Walter} F.,
  {Smail} I.,  {Dunlop} J.~S.,  {Knudsen} K.~K.,  {Alexander} D.~M.,
  {Bertoldi} F.,  {Brandt} W.~N.,  {Chapman} S.~C.,  {Cox} P.,  {Dannerbauer}
  H.,  {De Breuck} C.,  {Gawiser} E.,  {Ivison} R.~J.,    {Lutz} D. e.~a.,
  2009, ApJ, 707, 1201

\bibitem[\protect\citeauthoryear{{Werner}, {Roellig}, {Low}, {Rieke}, {Rieke},
  {Hoffmann}, {Young}, {Houck}, {Brandl}, {Fazio}, {Hora}, {Gehrz}, {Helou},
  {Soifer}, {Stauffer}, {Keene}, {Eisenhardt}, {Gallagher} \&
  {Gautier}}{{Werner} et~al.}{2004}]{werner_2004}
{Werner} M.~W.,  {Roellig} T.~L.,  {Low} F.~J.,  {Rieke} G.~H.,  {Rieke} M.,
  {Hoffmann} W.~F.,  {Young} E.,  {Houck} J.~R.,  {Brandl} B.,  {Fazio} G.~G.,
  {Hora} J.~L.,  {Gehrz} R.~D.,  {Helou} G.,  {Soifer} B.~T.,  {Stauffer} J.,
  {Keene} J.,  {Eisenhardt} P.,  {Gallagher} D.,    {Gautier} T.~N. e.~a.,
  2004, ApJS, 154, 1

\bibitem[\protect\citeauthoryear{{Wiebe}, {Ade}, {Bock}, {Chapin}, {Devlin},
  {Dicker}, {Griffin}, {Gundersen}, {Halpern}, {Hargrave}, {Hughes}, {Klein},
  {Marsden}, {Martin}, {Mauskopf}, {Netterfield}, {Olmi}, {Pascale} \&
  {Patanchon}}{{Wiebe} et~al.}{2009}]{wiebe_2009}
{Wiebe} D.~V.,  {Ade} P.~A.~R.,  {Bock} J.~J.,  {Chapin} E.~L.,  {Devlin}
  M.~J.,  {Dicker} S.,  {Griffin} M.,  {Gundersen} J.~O.,  {Halpern} M.,
  {Hargrave} P.~C.,  {Hughes} D.~H.,  {Klein} J.,  {Marsden} G.,  {Martin}
  P.~G.,  {Mauskopf} P.,  {Netterfield} C.~B.,  {Olmi} L.,  {Pascale} E.,
  {Patanchon} G. e.~a.,  2009, ApJ, 707, 1809

\bibitem[\protect\citeauthoryear{{Wilson}, {Austermann}, {Perera}, {Scott},
  {Ade}, {Bock}, {Glenn}, {Golwala}, {Kim}, {Kang}, {Lydon}, {Mauskopf},
  {Predmore}, {Roberts}, {Souccar} \& {Yun}}{{Wilson}
  et~al.}{2008}]{wilson_2008}
{Wilson} G.~W.,  {Austermann} J.~E.,  {Perera} T.~A.,  {Scott} K.~S.,  {Ade}
  P.~A.~R.,  {Bock} J.~J.,  {Glenn} J.,  {Golwala} S.~R.,  {Kim} S.,  {Kang}
  Y.,  {Lydon} D.,  {Mauskopf} P.~D.,  {Predmore} C.~R.,  {Roberts} C.~M.,
  {Souccar} K.,    {Yun} M.~S.,  2008, MNRAS, 386, 807

\bibitem[\protect\citeauthoryear{{Wilson}, {Ferris}, {Axtens}, {Brown},
  {Davis}, {Hampson}, {Leach}, {Roberts}, {Saunders}, {Koribalski}, {Caswell},
  {Lenc} \& {Stevens}}{{Wilson} et~al.}{2011}]{wilson_2011}
{Wilson} W.~E.,  {Ferris} R.~H.,  {Axtens} P.,  {Brown} A.,  {Davis} E.,
  {Hampson} G.,  {Leach} M.,  {Roberts} P.,  {Saunders} S.,  {Koribalski}
  B.~S.,  {Caswell} J.~L.,  {Lenc} E.,    {Stevens} J. e.~a.,  2011, MNRAS,
  416, 832

\bibitem[\protect\citeauthoryear{{Wright}, {Griffith}, {Burke} \&
  {Ekers}}{{Wright} et~al.}{1994}]{wright_1994}
{Wright} A.~E.,  {Griffith} M.~R.,  {Burke} B.~F.,    {Ekers} R.~D.,  1994,
  ApJS, 91, 111

\bibitem[\protect\citeauthoryear{{Wright}}{{Wright}}{1976}]{wright_1976}
{Wright} E.~L.,  1976, ApJ, 210, 250

\bibitem[\protect\citeauthoryear{{Wright}, {Chen}, {Odegard}, {Bennett},
  {Hill}, {Hinshaw}, {Jarosik}, {Komatsu}, {Nolta}, {Page}, {Spergel},
  {Weiland}, {Wollack}, {Dunkley}, {Gold}, {Halpern}, {Kogut}, {Larson},
  {Limon} \& {Meyer}}{{Wright} et~al.}{2009}]{wright_2009}
{Wright} E.~L.,  {Chen} X.,  {Odegard} N.,  {Bennett} C.~L.,  {Hill} R.~S.,
  {Hinshaw} G.,  {Jarosik} N.,  {Komatsu} E.,  {Nolta} M.~R.,  {Page} L.,
  {Spergel} D.~N.,  {Weiland} J.~L.,  {Wollack} E.,  {Dunkley} J.,  {Gold} B.,
  {Halpern} M.,  {Kogut} A.,  {Larson} D.,  {Limon} M.,    {Meyer} S.~S. e.~a.,
   2009, ApJS, 180, 283

\end{thebibliography}

\end{document}